\documentclass[aps,prd,preprintnumbers,superscriptaddress,nofootinbib,notitlepage,floatfix,preprint]{revtex4-2}
\usepackage[pdftex]{graphicx}
\usepackage{bm,latexsym,amsmath,amssymb,amsfonts,mathrsfs}
\usepackage{float}
\usepackage{color}
\usepackage{physics}
\usepackage{color}
\usepackage{comment}
\usepackage{here}
\usepackage{cancel}
\usepackage{empheq}
\usepackage{stmaryrd}
\allowdisplaybreaks[1]
\usepackage[pdftex,colorlinks=true,linkcolor=blue,citecolor=cyan,backref=page]{hyperref}
\makeatletter
\let\MYcaption\@makecaption
\makeatother

\usepackage{subcaption}
\makeatletter
\let\@makecaption\MYcaption
\makeatother

\renewcommand{\theequation}{\arabic{section}.\arabic{equation}}
\makeatletter
\@addtoreset{equation}{section}

\makeatother

\begin{document}
\title{Periapsis shift in the Zipoy-Voorhees spacetime}
%
\author{Akihito~Katsumata\footnote{Author to whom any correspondence should be addressed.}}
\email[Email: ]{a.katsumata@rikkyo.ac.jp}
\affiliation{Department of Physics, Rikkyo University, Toshima, Tokyo 171-8501, Japan}
\author{Tomohiro~Harada}
\email[Email: ]{harada@rikkyo.ac.jp}
\affiliation{Department of Physics, Rikkyo University, Toshima, Tokyo 171-8501, Japan}
\date{\today}
%
\begin{abstract}
We study the periapsis shift of timelike bound orbits in the equatorial plane in the Zipoy-Voorhees spacetime, which is an exact, static, axisymmetric, and vacuum solution characterized by the deformation parameter $\gamma$, including the Schwarzschild spacetime as $\gamma = 1$. We newly derive the formula for the periapsis shift by the post-Newtonian (PN) expansion and show that the quadrupole moment contributes to the periapsis shift apparently as in the 2PN order. Applying this formula to observational data on S2, a star orbiting closely to Sagittarius A* (Sgr A*), we show that the parameter $\gamma$ is constrained to $\gamma \gtrsim 1.9 \times 10^{-2}$ for the gravitational field of Sgr A*. This is equivalent to $\tilde{M}_2 \lesssim 9.7 \times 10^2$ in terms of the dimensionless quadrupole moment $\tilde{M}_2$, that is, the supermassive compact object at Sgr A* cannot be very prolate, whereas $\tilde{M}_2 \geq -1/3$ is required from the solution. Finally, we derive a new series representation in this spacetime using a recently proposed prescription, which shows fast convergence not only in the weak-field regime with not necessarily small eccentricity but also in the strong-field regime with small eccentricity.
\end{abstract}
\preprint{RUP-25-15}
\maketitle
\newpage
\tableofcontents
\section{Introduction} \label{sec:intro} 
Recent advances in observational techniques have greatly improved our ability to probe the nature of the supermassive compact object at Sagittarius A* (Sgr A*), a candidate for a black hole (BH). For example, the imaging of Sgr A* by the Event Horizon Telescope and discussions regarding it have been reported~\cite{EventHorizonTelescope:2022wkp,EventHorizonTelescope:2022apq,EventHorizonTelescope:2022wok,EventHorizonTelescope:2022exc,EventHorizonTelescope:2022urf,EventHorizonTelescope:2022xqj,EventHorizonTelescope:2024hpu,EventHorizonTelescope:2024rju,Vagnozzi:2022moj}. In addition, orbital motions of S-stars around Sgr A* have been also investigated~\cite{Ghez:2000ay,Schodel:2002py,Ghez:2003rt,Gillessen:2008qv,Ghez:2008ms,Saida:2024cjc}. They provide us with important insights into the nature of the spacetime geometry and matter distribution around Sgr A*. Therefore, a theoretical understanding of the orbital motions is crucial for interpreting and discussing the results of the observations.

One of the important relativistic effects in orbital motions is the periapsis shift, a phenomenon in which the major axis of the orbit rotates and the periapsis is shifted. The most famous example of this phenomenon is Einstein's successful explanation of Mercury's perihelion shift~\cite{Einstein:1916vd}. Recently, the periapsis shift of S2, which is a star orbiting around Sgr A*, has been observed by the Gravity collaboration~\cite{GRAVITY:2020gka}. Although Sgr A* is usually considered to be a Kerr BH, the true identity remains unclear. Motivated by this, various alternative possibilities have been investigated. For example, periapsis shifts around naked singularities~\cite{Bini:2005dy,Bambhaniya:2021ybs,Ota:2021mub,Bambhaniya:2025xmu}, in dark matter distribution around a BH~\cite{Igata:2022rcm}, and in dark matter distribution with a dense core~\cite{Igata:2022nkt} have been investigated. Additionally, general formulae for the periapsis shift of a quasi-circular orbit in static spherically symmetric spacetimes have been derived~\cite{Harada:2022uae}.

The Zipoy-Voorhees (ZV) spacetime~\cite{Bach:1922,Darmois:1927,Zipoy:1966btu,Voorhees:1970ywo}, sometimes referred to as the $\gamma$-metric, $\delta$-metric, and $q$-metric, has attracted the attention of many researchers as a candidate for the spacetime around Sgr A*. This spacetime is an exact, static, axisymmetric, asymptotically flat, and vacuum solution of Einstein's field equations, and belongs to the Weyl solutions~\cite{Weyl:1917,Weyl:2012}. The ZV spacetime is characterized by the deformation parameter $\gamma$, and for appropriate values of $\gamma$, the spacetime reproduces the Schwarzschild~\cite{Sch1916} and the Chazy-Curzon (CC)~\cite{Chazy:1924,Curzon:1925} spacetimes. Although the ZV spacetime has a naked singularity, it may be physically meaningful as an external solution. The global structure of the ZV spacetime is studied in~\cite{Kodama:2003ch}. In~\cite{Abdikamalov:2019ztb}, the possibility that the source of the ZV spacetime behaves as a so-called BH mimicker has been discussed by examining the lensing effect and shadow. In~\cite{Lora-Clavijo:2023ukh}, it has been shown that the ZV naked singularity provides a good fit to the observed astrometric and spectroscopic data of the S2 star, similar to the Schwarzschild BH. Also, removing the singularity from the ZV spacetime by a thin shell has been discussed in~\cite{Saito:2024hzc}. The geodesic motions of massive and massless particles in the ZV spacetime have been investigated in e.g.~\cite{Herrera:1998rj,Chowdhury:2011aa,Boshkayev:2015jaa,Abdikamalov:2019ztb,Momynov:2024bhn,Idrissov:2025ugs}. Furthermore, the relativistic effects such as the Shapiro time delay in the ZV spacetime and their constraints have been discussed in~\cite{Chakrabarty:2022fbd,Utepova:2025jkx}.

Recently, a formula for the periapsis shift in the ZV spacetime has been derived in the approximation that the eccentricity of the orbit, $e$, is sufficiently small, and using that and the observational data on the periapsis shift of S2~\cite{GRAVITY:2020gka}, the deformation parameter $\gamma$ of Sgr A* has been constrained~\cite{Chakrabarty:2022fbd}. However, since the eccentricity of S2 is $e \simeq 0.88$~\cite{GRAVITY:2020gka}, it is inappropriate to apply the formula derived under the approximation. Therefore, the constraint on $\gamma$ obtained by the analysis is not correct. Indeed, in~\cite{Chakrabarty:2022fbd}, the authors note that their study is preliminary. A more accurate analysis for S2-Sgr A* would require either a post-Newtonian (PN) expansion formula for arbitrary eccentricities or numerical simulations as mentioned by the authors in~\cite{Chakrabarty:2022fbd}.

Motivated by the above, in this paper, we derive the PN expansion formula for the periapsis shift in the ZV spacetime with arbitrary eccentricity and compare it with that in the Schwarzschild spacetime. Note that due to the non-spherical symmetry of the spacetime, orbits in the ZV spacetime are not generally confined to the equatorial plane and would cause, for example, precession of the orbital plane in addition to the periapsis shift. Nevertheless, as a first step, this paper focuses on orbits confined to the equatorial plane. We also give constraints on the deformation parameter $\gamma$ and the corresponding quadrupole moment $\tilde{M}_2$ of Sgr A* using the obtained PN expansion formula and the observational data on the periapsis of S2~\cite{GRAVITY:2020gka}. In this analysis, we again assume that the orbital plane of S2 is equatorial. Note that for a more realistic analysis, we should consider cases where the orbit is inclined relative to the equatorial plane.

For the analysis of the periapsis shift in the S2-Sgr A* system, the PN expansion formula for arbitrary eccentricity is sufficiently accurate. However, a star orbiting closer to Sgr A* than S2 has been reported recently~\cite{Peißker_2022}, and when the periapsis shift of such a star in strong gravitational fields is observed in the future, the PN expansion formula may not be sufficiently good. Therefore, in the last part of this paper, we derive a new series representation for the periapsis shift in the ZV spacetime using a recently proposed prescription~\cite{Katsumata:2024qzv}. In this prescription, the expansion parameter is defined as the eccentricity divided by a dimensionless quantity that becomes zero in the limit of the innermost stable circular orbit (ISCO). The obtained formula allows us to easily and very accurately calculate the periapsis shift in the strong-field regime where the convergence of the PN expansion formula is not guaranteed, and will also be useful for verifying numerical calculations.

This paper is organized as follows. In Sec.~\ref{sec:peri_PN}, we consider the geodesic motion of massive test particles in the ZV spacetime and derive the PN expansion formula for the periapsis shift with arbitrary eccentricity in this spacetime. We also discuss the deviation of the obtained PN expansion formula from that in the Schwarzschild spacetime. In Sec.~\ref{sec:constraint}, we give constraints to the deformation parameter $\gamma$ and the corresponding quadrupole moment $\tilde{M}_2$ of Sgr A* using the obtained PN expansion formula and the observational data on the periapsis shift of S2~\cite{GRAVITY:2020gka}. In Sec.~\ref{sec:another_series}, we derive a new series representation for the periapsis shift in the ZV spacetime using the recently proposed prescription~\cite{Katsumata:2024qzv}. The behaviors of the new expansion parameter and the relative errors of the new representation are also examined. Sec.~\ref{sec:sum_conc} is devoted to the summary and conclusions. In Appendix~\ref{ZV_PN_high_order}, we present the PN expansion formula for the periapsis shift in the ZV spacetime with higher PN order terms. In Appendix~\ref{app:paras}, we present the orbital parameters of S2-Sgr A* that we use to constrain the deformation of Sgr A* in Sec.~\ref{sec:constraint}. In Appendix~\ref{app:ps_qc}, we derive an exact formula for the periapsis shift of a quasi-circular orbit in the ZV spacetime using a simple method, different from the PN expansion. In Appendix~\ref{Yn_forms}, the concrete forms of expansion coefficients for the new series expansion, which are omitted in the main text, are explicitly shown. In Appendices~\ref{app:review_ZV} and~\ref{app:multipole_moments}, a brief review of the ZV spacetime and a calculation of the relativistic multipole moments in this spacetime are given, respectively. Throughout this paper, the geometrical units with $G=c=1$ and the sign convention for the metric $(-,+,+,+)$ are used unless otherwise stated.

\section{PN expansion formula for the periapsis shift} \label{sec:peri_PN}
In this section, we consider timelike bound orbits in the ZV spacetime and aim to derive the PN expansion formula for the periapsis shift in this spacetime. We also discuss the deviation of the PN expansion formula from the spherically symmetric case.
\subsection{Motion of massive test particles}
First, we briefly introduce the ZV spacetime and consider the timelike geodesics in this spacetime as a preliminary for deriving the PN expansion formula for the periapsis shift. The ZV spacetime, also known as the $\gamma$-metric, $\delta$-metric, and $q$-metric, is an exact, static, axisymmetric, asymptotically flat, and vacuum solution of Einstein's field equations. In the Erez-Rosen coordinates~\cite{Erez:1959}, the line element is given by 
\begin{align}
    \dd{s}^2 = -f^\gamma \dd{t}^2 + f^{\gamma(\gamma-1)} g^{1-\gamma^2} \left( f^{-1} \dd{r}^2 + r^2 \dd{\theta}^2 \right) + f^{1-\gamma} r^2 \sin^2 \theta \dd{\phi}^2, \label{ZV_met_ER}
\end{align}
where
\begin{align}
    f = 1 - \frac{2M}{\gamma r}, \quad g = 1 - \frac{2M}{\gamma r} + \frac{M^2 \sin^2 \theta}{\gamma^2 r^2}.
\end{align}
The parameter $M$ denotes the Arnowitt-Deser-Misner (ADM) mass~\cite{Arnowitt:1960es}. The dimensionless parameter $\gamma$ is often called the deformation parameter, which is related to the mass multipole moments and characterizes the deformation of the source from spherical symmetry. Now, we can assume $\gamma>0$ without loss of generality (see Appendix~\ref{app:review_ZV}). As for $\gamma >1$ ($0<\gamma<1$), the spacetime is oblate (prolate). The spacetime is spherically symmetric for $\gamma = 1$, and it recovers the Schwarzschild spacetime~\cite{Sch1916}. In this paper, we will refer to the radial coordinate in the obtained standard form of the Schwarzschild metric as the ``\textit{Schwarzschild radial coordinate}". Note that the spacetime also recovers the CC spacetime~\cite{Chazy:1924, Curzon:1925} in spherical coordinates in the limit of $\gamma \rightarrow \infty$. A brief review of the ZV spacetime and a calculation of relativistic multipole moments in the spacetime are given in Appendices~\ref{app:review_ZV} and~\ref{app:multipole_moments}, respectively.

Let us consider the geodesic motion of a test particle with mass $\mu$ on the equatorial plane (i.e., $\theta = \pi/2$). The Lagrangian of the particle is written as
\begin{align}
    \mathcal{L} = \frac{1}{2} \mu \left( -f^\gamma \dot{t}^2 + f^{\gamma^2 - \gamma -1} g^{1-\gamma^2} \dot{r}^2 + f^{1-\gamma} r^2 \dot{\phi}^2 \right), \label{Lag}
\end{align}
where the dot denotes the derivative with respect to the proper time of the particle. Due to the spacetime symmetry, the energy and the angular momentum of the particle
\begin{align}
    E := - \pdv{\mathcal{L}}{\dot{t}}, \quad L := \pdv{\mathcal{L}}{\dot{\phi}},
\end{align}
are conserved. Evaluating these derivatives from Eq.~\eqref{Lag} and solving the two equations for $\dot{t}$ and $\dot{\phi}$, we obtain
\begin{align}
    \dot{t} = \frac{\tilde{E}}{f^\gamma}, \quad \dot{\phi} = \frac{\tilde{L}}{r^2 f^{1-\gamma}}, \label{tdot_phidot}
\end{align}
where $\tilde{E} := E/\mu$ and $\tilde{L} := L/\mu$. Substituting Eq.~\eqref{tdot_phidot} into the normalization condition $g_{\mu \nu} \dot{x}^\mu \dot{x}^\nu = -1$, we obtain after some simplification
\begin{align}
    \dot{r}^2 = \left( \frac{f}{g} \right)^{1-\gamma^2} \left[ \tilde{E}^2 - f^{\gamma} \left( 1 + f^{\gamma-1} \frac{\tilde{L}^2}{r^2} \right) \right]. \label{rdot}
\end{align}

\subsection{Derivation} \label{subsec:devi}
In this subsection, we derive the PN expansion formula for the periapsis shift in the ZV spacetime with arbitrary eccentricity. Since $\dot{r}=0$ at the periapsis $r=r_p$ and the apoapsis $r=r_a$, we can get two equations from Eq.~\eqref{rdot} as
\begin{align}
    0 = \tilde{E}^2 - f_p^{\gamma} \left( 1 + f_p^{\gamma-1} \frac{\tilde{L}^2}{r_p^2} \right), \quad 0 = \tilde{E}^2 - f_a^{\gamma} \left( 1 + f_a^{\gamma-1} \frac{\tilde{L}^2}{r_a^2} \right),
\end{align}
where $f_p := f(r_p) = 1-2M/({\gamma}r_p)$ and $f_a := f(r_a) = 1-2M/({\gamma}r_a)$. Solving these equations for $\tilde{E}^2$ and $\tilde{L}^2$, we obtain
\begin{align}
    \tilde{E}^2 = \frac{(f_a f_p)^\gamma \left( r_a^2 f_p^{\gamma -1} - r_p^2 f_a^{\gamma-1} \right)}{ r_a^2 f_p^{2\gamma-1} - r_p^2 f_a^{2 \gamma-1} }, \quad \tilde{L}^2 = \frac{r_a^2 r_p^2 \left(f_a^{\gamma }-f_p^{\gamma }\right)}{r_a^2 f_p^{2 \gamma-1}-r_p^2 f_a^{2 \gamma-1}}. \label{eq:E2_and_L2}
\end{align}
From Eqs.~\eqref{tdot_phidot} and \eqref{rdot}, one can obtain the change in $\phi$ as the particle moves from the periapsis to the next periapsis as
\begin{align}
    \delta \phi_\text{ZV} &:= 2 \int_{r_p}^{r_a} \left| \dv{\phi}{r} \right| \dd{r} \label{eq:peri_shift_def} \\[8pt]
    &= 2 \int_{r_p}^{r_a} \frac{\tilde{L}}{r^2 f^{1-\gamma}} \left\{ \left( \frac{f}{g} \right)^{1-\gamma^2} \left[ \tilde{E}^2 - f^{\gamma} \left( 1 + f^{\gamma-1} \frac{\tilde{L}^2}{r^2} \right) \right]  \right\}^{-1/2} \dd{r}. \label{peri_seki1}
\end{align}
The periapsis shift per round is defined by subtracting the contribution of Newtonian gravity $2\pi$ from $\delta \phi$:
\begin{align}
    \Delta \phi_\text{ZV} := \delta \phi_\text{ZV} - 2 \pi. \label{peri_ZV_teigi}
\end{align}

\begin{figure}[htbp]
  \begin{minipage}{0.325\linewidth}
    \centering
    \includegraphics[keepaspectratio, scale=0.585]{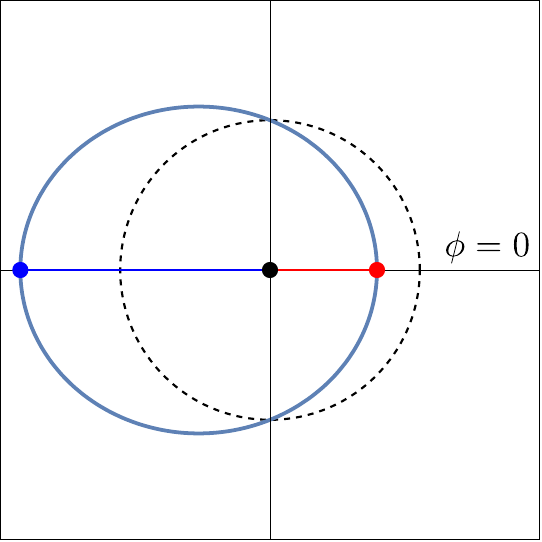}
    \subcaption{Newtonian case}
    \label{fig:peri_qc_newton}
  \end{minipage} 
  \begin{minipage}{0.325\linewidth}
    \centering
    \includegraphics[keepaspectratio, scale=0.585]{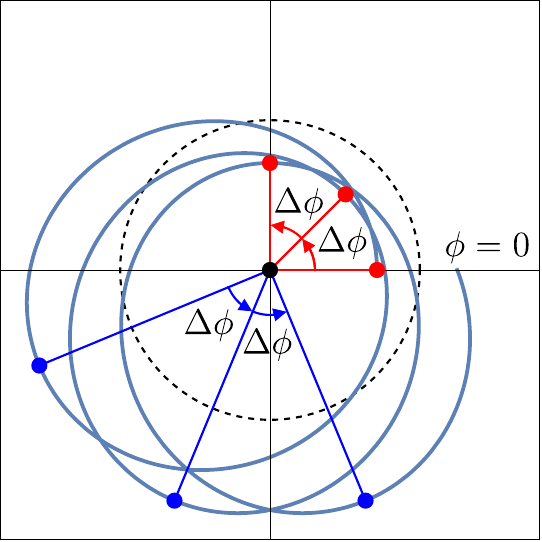}
    \subcaption{Forward shift case}
    \label{fig:peri_qc_pro}
  \end{minipage}
  \begin{minipage}{0.325\linewidth}
    \centering
    \includegraphics[keepaspectratio, scale=0.585]{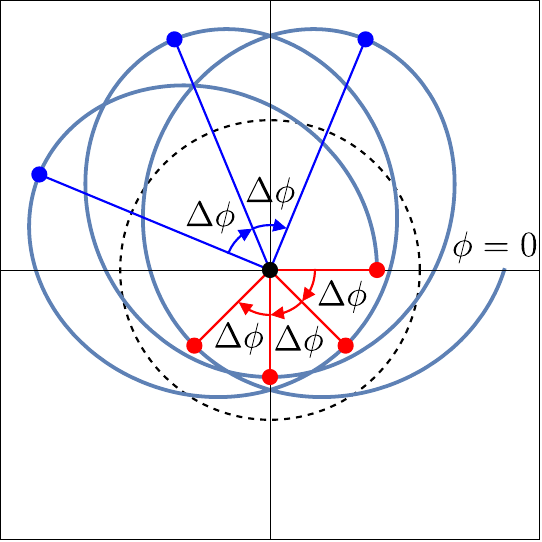}
    \subcaption{Backward shift case}
    \label{fig:peri_qc_ret}
  \end{minipage}
  \caption{The schematic illustration of the periapsis shift. The dashed black line, the black dot, the red dot, and the blue dot represent the circular orbit with zero eccentricity, the position of the gravitational source, the periapsis, and the apoapsis, respectively. The orbits start from the periapsis at $\phi=0$ and evolve counterclockwise. In the Newtonian case (Fig.~\ref{fig:peri_qc_newton}), the periapsis is not shifted. Fig.~\ref{fig:peri_qc_pro} shows the forward shift case, in which the periapsis is shifted in the same direction as the orbital motion, with a shift angle $\Delta \phi$. In contrast, Fig.~\ref{fig:peri_qc_ret} shows the backward shift case, in which the periapsis is shifted in the opposite direction to the orbital motion. Note that the shift can be very large in the strong-field regime even for an infinitesimally small eccentricity.}
  \label{fig:peri_qc}
\end{figure}

We present a schematic illustration of the periapsis shift in Fig.~\ref{fig:peri_qc}. The dashed black line, the black dot, the red dot, and the blue dot represent the circular orbit with zero eccentricity, the position of the gravitational source, the periapsis, and the apoapsis, respectively. The orbits start from the periapsis at $\phi=0$ and evolve counterclockwise. Fig.~\ref{fig:peri_qc_newton} shows the Newtonian case. In this case, the periapsis is not shifted since the orbit closes in one revolution. Fig.~\ref{fig:peri_qc_pro} shows the case in which the orbit completes a revolution before reaching the next periapsis, and the periapsis is shifted in the same direction as the orbital motion (forward shift), with a shift angle $\Delta \phi$. Conversely, Fig.~\ref{fig:peri_qc_ret} shows the case in which the orbit completes a revolution after reaching the next periapsis, and the periapsis is shifted in the opposite direction to the orbital motion (backward shift). It should be emphasized that Fig.~\ref{fig:peri_qc} is intended only as a schematic representation, in which, for example, the eccentricity and the periapsis shift are exaggerated for clarity. Note that, as we will discuss in detail in Appendix~\ref{subsec:qc}, the same qualitative discussion can also be applied to the quasi-circular orbits, which have infinitesimally small eccentricity.

Here, let us define the eccentricity, the semi-major axis, and the semi-latus rectum as
\begin{align}
    \bar{e} := \frac{r_a-r_p}{r_a+r_p}, \quad \bar{d} := \frac{r_p+r_a}{2}, \quad \bar{p} := \frac{2 r_p r_a}{r_p + r_a} = \bar{d} (1-\bar{e}^2), \label{ER_e_d_p}
\end{align}
respectively. Using Eq.~\eqref{ER_e_d_p}, $r_p$ and $r_a$ can be written as
\begin{align}
    r_p = \bar{d}(1-\bar{e}) = \frac{\bar{p}}{1+\bar{e}}, \quad r_a = \bar{d}(1+\bar{e}) = \frac{\bar{p}}{1-\bar{e}}. \label{eq:rp_ra_e_p}
\end{align}
We have used the bar to represent that the quantities are defined in terms of the Erez-Rosen radial coordinates, to distinguish them from those defined in terms of the circumferential radius, which we will introduce later. Note that these orbital parameters coincide with those defined in terms of the Schwarzschild and Weyl radial coordinate for $\gamma=1$ and $\gamma \rightarrow \infty$, respectively. This is because the Erez-Rosen radial coordinate coincides with the Schwarzschild and Weyl radial coordinates for $\gamma=1$ and $\gamma \rightarrow \infty$, respectively (see also Appendix~\ref{app:review_ZV}).

Introducing the variable transformation $1/r = (1+\bar{e} \sin \chi)/\bar{p}$, Eq.~\eqref{peri_seki1} becomes
\begin{align}
    \delta \phi_\text{ZV} = 2 \int_{-\pi/2}^{\pi/2} F(\chi) \dd{\chi}, \label{peri_seki2}
\end{align}
where the integrand is defined as
\begin{align}
    F(\chi) := \frac{\tilde{L} (\bar{e}/\bar{p}) \cos \chi }{f^{1-\gamma}} \left\llbracket \left( \frac{f}{g} \right)^{1-\gamma^2} \left\{ \tilde{E}^2 - f^{\gamma} \left[ 1 + f^{\gamma-1} \tilde{L}^2 \frac{(1+\bar{e} \sin \chi)^2}{\bar{p}^2} \right] \right\}  \right\rrbracket^{-1/2}. \label{eq:F_chi}
\end{align}
Substituting $\tilde{E}^2$ and $\tilde{L}^2$ given by Eq.~\eqref{eq:E2_and_L2}, where $r_p$ and $r_a$ are rewritten in terms of $\bar{e}$ and $\bar{p}$ by using Eq.~\eqref{eq:rp_ra_e_p}, into Eq.~\eqref{eq:F_chi}, we expand the resulting $F(\chi)$ in powers of $M/\bar{p}$. Putting the expanded $F(\chi)$ back in Eq.~\eqref{peri_seki2}, integrating it, and combining it with Eq.~\eqref{peri_ZV_teigi}, we obtain the PN expansion formula for the periapsis shift per round in the ZV spacetime as
\begin{align}
    \Delta \phi_\text{ZV} &= \bar{\Delta} \phi_\text{ZV,1PN} + \bar{\Delta} \phi_\text{ZV,2PN} + \cdots, \label{ZV_peri_PN}
\end{align}
where
\begin{align}
    \bar{\Delta} \phi_\text{ZV,1PN} := \frac{6 \pi M}{\bar{p}}, \quad \bar{\Delta} \phi_\text{ZV,2PN} := \left[ \left(44+\frac{12}{\gamma}-\frac{2}{\gamma ^2}\right)+3\left(-3+\frac{4}{\gamma }\right) \bar{e}^2 \right] \frac{\pi M^2}{2 \bar{p}^2}. \label{ZV_peri_PN_v2}
\end{align}
The subscript label $n$PN ($n=1,2,3,\ldots$) on each term in Eqs.~\eqref{ZV_peri_PN} and~\eqref{ZV_peri_PN_v2} denotes the order of the PN expansion. The formula with even higher PN order terms is presented in Appendix~\ref{ZV_PN_high_order}. It should be noted that Eq.~\eqref{ZV_peri_PN} is expressed in terms of the Erez-Rosen radial coordinate $r$, not in the Schwarzschild or Weyl radial coordinates used in e.g.~\cite{Bini:2005dy,Katsumata:2024qzv}.

Note that Eq.~\eqref{ZV_peri_PN} recovers the PN expansion formula in the Schwarzschild spacetime~\cite{Einstein:1916vd} (see also e.g.~\cite{poisson_will_2014,Tucker:2018rgy,He:2023joa}) for $\gamma=1$:
\begin{align}
    \Delta \phi_\text{S} = \frac{6 \pi M}{\bar{p}}  +\frac{3 \pi (18+\bar{e}^2) M^2}{2 \bar{p}^2} + \cdots,
    \label{eq:PN_S}
\end{align}
and in the CC spacetime~\cite{Bini:2005dy,Katsumata:2024qzv} for $\gamma \rightarrow \infty$:
\begin{align}
    \Delta \phi_\text{CC} = \frac{6 \pi M}{\bar{p}} + \frac{\pi (44 - 9 \bar{e}^2) M^2}{2 \bar{p}^2} + \cdots. \label{eq:PN_CC}
\end{align}
As mentioned below Eq.~\eqref{ER_e_d_p}, this is because the Erez-Rosen radial coordinate coincides with the Schwarzschild and Weyl radial coordinates for $\gamma=1$ and $\gamma \rightarrow \infty$, respectively (see also Appendix~\ref{app:review_ZV}), and therefore the definitions of the eccentricity and the semi-latus rectum coincide. We can also find that the 1PN order term of the PN expansion formula \eqref{ZV_peri_PN} does not depend on $\gamma$, and is the same as those in the Schwarzschild and the CC spacetimes. 

\subsection{Deviation from the Schwarzschild spacetime}
Let us discuss the deviation of the PN expansion formula for the periapsis shift in the ZV spacetime~\eqref{ZV_peri_PN} from that in the Schwarzschild spacetime. It is important to note here that the eccentricity $\bar{e}$ and the semi-latus rectum $\bar{p}$ in Eq.~\eqref{ZV_peri_PN} are defined in terms of the Erez-Rosen radial coordinate. It will be appropriate to rewrite these quantities in terms of the circumferential radius to compare them. From the line element~\eqref{ZV_met_ER}, the circumferential radius $R$ on the equatorial plane can be calculated as
\begin{align}
    R = \frac{1}{2\pi} \int_0^{2\pi} f^{(1-\gamma)/2} r \dd{\phi} = r \left(1 - \frac{2M}{\gamma r}\right)^{(1-\gamma)/2}. \label{ER_to_CR_exact}
\end{align}
Note that, in the Schwarzschild case ($\gamma=1$), we find $R=r$, which means the circumferential radius coincides with the Schwarzschild radial coordinate.

From Eq.~\eqref{ER_to_CR_exact}, if $M/r \ll 1$, we obtain the following approximate formula, which rewrites the Erez-Rosen radial coordinate $r$ in terms of the circumferential radius $R$:
\begin{align}
    r \simeq R - \left(1-\frac{1}{\gamma}\right) M. \label{ER_to_CR}
\end{align}
Here, we define the eccentricity, the semi-major axis, and the semi-latus rectum in terms of the circumferential radius, as in Eq.~\eqref{ER_e_d_p}:
\begin{align}
    e := \frac{R_a-R_p}{R_a+R_p}, \quad d := \frac{R_p+R_a}{2}, \quad p := \frac{2 R_p R_a}{R_p + R_a} = d (1-e^2),
\end{align}
respectively, where $R_p$ and $R_a$ are the periapsis and apoapsis distances, respectively. Using these quantities, $R_p$ and $R_a$ can be written as
\begin{align}
    R_p = d(1-e) = \frac{p}{1+e}, \quad R_a = d(1+e) = \frac{p}{1-e}. \label{CR_Rp_Ra}
\end{align}
From Eqs.~\eqref{ER_e_d_p}, \eqref{ER_to_CR}, and~\eqref{CR_Rp_Ra}, we obtain the formulae that transform the eccentricity and the semi-latus rectum (with bars) defined in terms of the Erez-Rosen radial coordinate into those defined in terms of the circumferential radius:
\begin{align}
    \bar{e} = e \left[ 1 + \mathcal{O}(M/p) \right], \quad \frac{M}{\bar{p}} = \frac{M}{p}
    \left\{ 1 + \frac{(1+e^2)(1 -1/\gamma) M}{p} + \mathcal{O}[(M/p)^2] \right\}.
\end{align}
Using these transformations, Eq.~\eqref{ZV_peri_PN} is rewritten as 
\begin{align}
    \Delta \phi_\text{ZV} \simeq \Delta \phi_\text{ZV,1PN} + \Delta \phi_\text{ZV,2PN}, \label{eq:ZV_PN_CR}
\end{align}
where 
\begin{align}
    \Delta \phi_\text{ZV,1PN} := \frac{6 \pi M}{p}, \quad \Delta \phi_\text{ZV,2PN} := \frac{\pi(56 +3 e^2 -2/\gamma^2)M^2}{2p^2}.
\end{align}
This is the PN expansion formula for the periapsis shift in the ZV spacetime in terms of the circumferential radius. In the Schwarzschild case ($\gamma=1$), Eq.~\eqref{eq:ZV_PN_CR} reduces to the same form as Eq.~\eqref{eq:PN_S}:
\begin{align}
    \Delta \phi_\text{S} \simeq \Delta \phi_\text{S,1PN} + \Delta \phi_\text{S,2PN}, \label{eq:Sch_PN_CR}
\end{align}
where
\begin{align}
    \Delta \phi_\text{S,1PN} := \frac{6 \pi M}{p}, \quad \Delta \phi_\text{S,2PN} := \frac{3 \pi (18 + e^2 ) M^2}{2 p^2}. \label{eq:Sch_PN_CR_2}
\end{align}
This is also consistent with Eq.~\eqref{eq:PN_S} since $\bar{e}=e$ and $\bar{p}=p$ for $\gamma=1$ ($R=r$).

Furthermore, let us define the dimensionless quadrupole moment as a more physically understandable parameter than $\gamma$ as follows:
\begin{align}
    \tilde{M}_2 := \frac{M_2}{M_0^3} = -\frac{1}{3} \left(1-\frac{1}{\gamma^2} \right) \quad \left( \tilde{M}_2 \geq -\frac{1}{3} \right), \label{eq:nd_qm}
\end{align}
where $M_0$ and $M_2$ are the monopole and quadrupole moments, respectively, and their concrete forms are given in Eqs.~\eqref{ZV_monopole} and \eqref{ZV_quadrupole} in Appendix~\ref{app:multipole_moments}. Note that $\tilde{M}_2 = 0$ for $\gamma = 1$ (Schwarzschild case) and $\tilde{M}_2 \rightarrow -1/3$ for $\gamma \rightarrow \infty$ (CC limit). It should also be noted that $\tilde{M}_2<0$ ($\tilde{M}_2>0$) corresponds to oblate (prolate) spacetime (see e.g.~\cite{Bianchi:2020bxa}). Rewriting $1/\gamma^2$ in Eq.~\eqref{eq:ZV_PN_CR} in terms of $\tilde{M}_2$ using Eq.~\eqref{eq:nd_qm}, we have
\begin{align}
    \Delta \phi_\text{ZV} \simeq \Delta \phi_\text{S} + \frac{\pi (1-1/\gamma^2) M^2}{p^2} = \Delta \phi_\text{S} - \frac{3 \pi M^2}{p^2} \tilde{M}_2. \label{eq:ZV_PN_CR_M2}
\end{align}
It can be seen that, in this formalism, the effect of the quadrupole moment apparently contributes to the periapsis shift in the 2PN order as the deviation from the spherically symmetric case. We can also find that the oblateness ($\tilde{M}_2 < 0$) and prolateness ($\tilde{M}_2 > 0$) contribute to increasing and decreasing the amount of the periapsis shift, respectively. This is consistent with known results in Newtonian gravity (see e.g.~\cite{Will:2018,Igata:2025trj}).

\begin{figure}[htbp]
\hspace{-10truemm}
  \begin{minipage}{0.45\linewidth}
    \centering
    \includegraphics[keepaspectratio, scale=0.9]{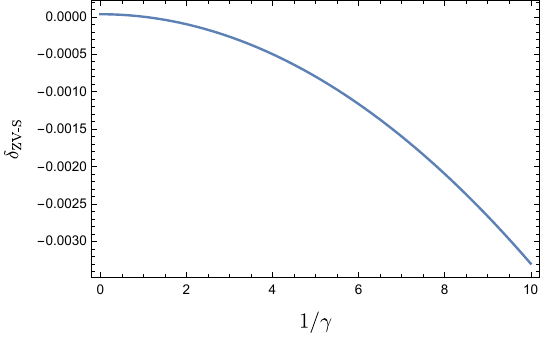}
    \subcaption{Plot with $1/\gamma$ on the horizontal axis}
    \label{fig:plot_pn_diff_gamma}
  \end{minipage} 
  \hspace{8truemm}
  \begin{minipage}{0.45\linewidth}
    \centering
    \includegraphics[keepaspectratio, scale=0.9]{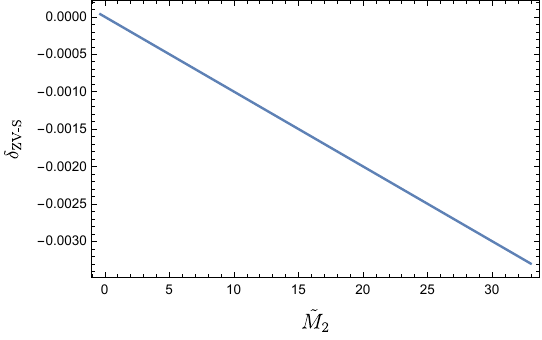}
    \subcaption{Plot with $\tilde{M}_2$ on the horizontal axis}
    \label{fig:plot_pn_diff_qm}
  \end{minipage}
  \\[5truemm]
  \hspace{-10truemm}
  \begin{minipage}{0.45\linewidth}
    \centering
    \includegraphics[keepaspectratio, scale=0.9]{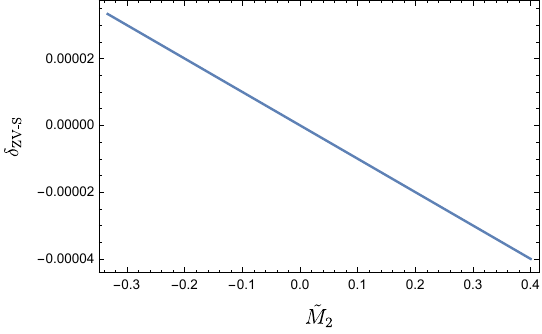}
    \subcaption{Enlarged plot of Fig.~\ref{fig:plot_pn_diff_qm}}
    \label{fig:plot_pn_diff_expand}
  \end{minipage}
  \caption{The plots of $\delta_\text{ZV-S}$, which is the deviation of the PN expansion formula for the periapsis shift in the ZV spacetime from that in the Schwarzschild spacetime. The horizontal axes are $1/\gamma$ (Fig.~\ref{fig:plot_pn_diff_gamma}) and $\tilde{M}_2$ (Fig.~\ref{fig:plot_pn_diff_qm}), and Fig.~\ref{fig:plot_pn_diff_expand} shows the enlarged view of Fig.~\ref{fig:plot_pn_diff_qm} near $\tilde{M}_2=0$. The eccentricity and the semi-latus rectum are fixed as $e=1/2$ and $M/p=1/5000$, respectively, as an example.}
  \label{fig:plot_pn_diff}
\end{figure}

Now, let us define the deviation of the PN expansion formula for the periapsis shift in the ZV spacetime from that in the Schwarzschild spacetime up to the 2PN order as
\begin{align}
    \delta_\text{ZV-S} &:= \frac{\left( \Delta \phi_\text{ZV,1PN} + \Delta \phi_\text{ZV,2PN} \right) - \left( \Delta \phi_\text{S,1PN} + \Delta \phi_\text{S,2PN} \right) }{ \Delta \phi_\text{S,1PN} + \Delta \phi_\text{S,2PN} } \\[8pt]
    &= \frac{2 \left( 1 - 1/\gamma^2 \right) M/p }{3 \left[ 4 + ( 18 + e^2) M/p \right]} = -\frac{2 M/p }{4 + (18+e^2) M/p} \tilde{M}_2.
    \label{devi_ZV_S}
\end{align}
We show the plots of the deviation $\delta_\text{ZV-S}$ in Fig.~\ref{fig:plot_pn_diff}. Figs.~\ref{fig:plot_pn_diff_gamma} and~\ref{fig:plot_pn_diff_qm} are with $1/\gamma$ and $\tilde{M}_2$ on the horizontal axis, respectively. Fig.~\ref{fig:plot_pn_diff_expand} is an enlarged plot of the region around $\tilde{M}_2=0$ in Fig.~\ref{fig:plot_pn_diff_qm}. From Eq.~\eqref{devi_ZV_S}, it can be seen that the deviation $\delta_\text{ZV-S}$ is proportional to $1/\gamma^2$ or $\tilde{M}_2$, and has a maximum value at $\gamma \rightarrow \infty$ ($\tilde{M}_2 \rightarrow -1/3$), which corresponds to the CC limit. Figs.~\ref{fig:plot_pn_diff_gamma} and~\ref{fig:plot_pn_diff_expand} shows this clearly.

\section{Constraint on the deformation parameter of Sgr A*} \label{sec:constraint}
In this section, we aim to give a constraint on the deformation parameter $\gamma$ of Sgr A* by using the obtained PN expansion formula~\eqref{eq:ZV_PN_CR}. The periapsis shift of S2 has been observed by the Gravity Collaboration~\cite{GRAVITY:2020gka} as
\begin{align}
    \Delta \phi_\text{S2} = f_\text{SP} \frac{6 \pi M}{p}, \quad f_\text{SP} = 1.1 \pm 0.19, \label{eq_peri_S2}
\end{align}
where $f_\text{SP}$ is the dimensionless factor which denotes the deviation from the 1PN order term of the PN expansion formula for the periapsis shift in the Schwarzschild spacetime [see Eq.~\eqref{eq:Sch_PN_CR_2}]. Assuming that the deviation $f_\text{SP}$ is a contribution from the 2PN order of the PN expansion formula in the ZV spacetime, we can rewrite Eq.~\eqref{eq_peri_S2} as
\begin{align}
    \frac{6 \pi M}{p} + \frac{\pi(56 +3 e^2 -2/\gamma^2)M^2}{2p^2} = f_\text{SP} \frac{6 \pi M}{p}.
\end{align}
Solving this equation for $f_\text{SP}$, we obtain
\begin{align}
    f_\text{SP} = 1+ \frac{(56 + 3 e^2 - 2 /\gamma^2) M}{12 p} = 1 + \frac{(18 + e^2-2 \tilde{M}_2) M}{4 p}. \label{func_f_gamma}
\end{align}
It can be found that $f_\text{SP}$ has a maximum value 
\begin{align}
    f_\text{SP,max} = 1 + \frac{(56 + 3 e^2) M}{12 p}, \label{f_max}
\end{align}
for $\gamma \rightarrow \infty$ ($\tilde{M}_2 \rightarrow -1/3$), which corresponds to the CC limit.

\begin{figure}[htbp]
\hspace{-10truemm}
  \begin{minipage}{0.45\linewidth}
    \centering
    \includegraphics[keepaspectratio, scale=0.9]{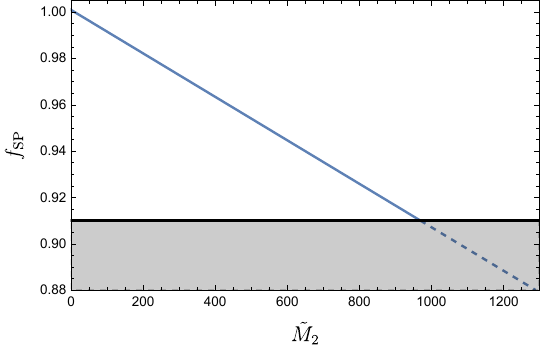}
    \subcaption{Overall}
    \label{fig:plot_f_overall}
  \end{minipage} 
  \hspace{8truemm}
  \begin{minipage}{0.45\linewidth}
    \centering
    \includegraphics[keepaspectratio, scale=0.9]{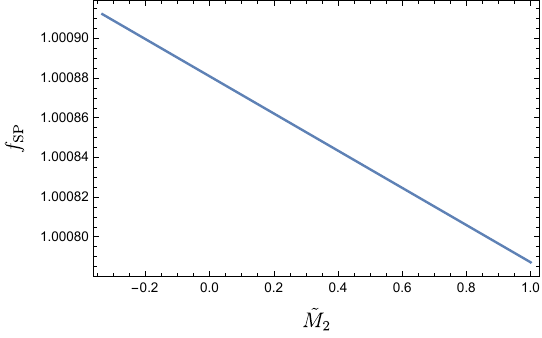}
    \subcaption{Near maximum value in Fig.~\ref{fig:plot_f_overall}}
    \label{fig:plot_f_nmv}
  \end{minipage}
  \\[5truemm]
  \hspace{-10truemm}
  \begin{minipage}{0.45\linewidth}
    \centering
    \includegraphics[keepaspectratio, scale=0.9]{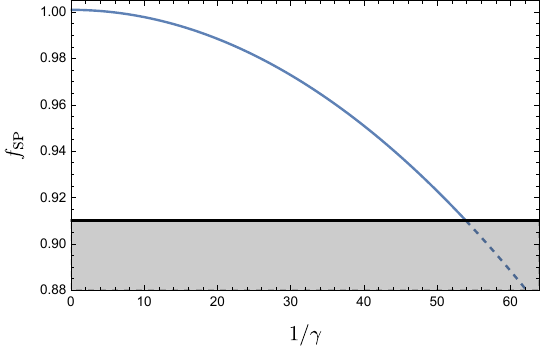}
    \subcaption{Plot with $1/\gamma$ on the horizontal axis}
    \label{fig:plot_f_gamma}
  \end{minipage} 
  \caption{The plots of $f_\text{SP}$ for S2-Sgr A*. Fig.~\ref{fig:plot_f_overall} shows an overall plot of $f_\text{SP}$. Fig.~\ref{fig:plot_f_nmv} shows an enlarged plot of Fig.~\ref{fig:plot_f_overall} near the maximum value. Fig.~\ref{fig:plot_f_gamma} is the plot with $1/\gamma$ as the horizontal axis. The region $f_\text{SP}<0.91$, which is restricted by the observation, is filled in grey. From Figs.~\ref{fig:plot_f_overall} and~\ref{fig:plot_f_gamma}, the observational constraint $f_\text{SP} \gtrsim 0.91$ allows us to constrain a lower bound on the value of $\gamma$ as $\gamma \gtrsim 1.9 \times 10^{-2}$, which corresponds to $\tilde{M}_2 \lesssim 9.7 \times 10^2$. Furthermore, if future observations find that $f_\text{SP}>1.00091$, we can exclude the possibility that the spacetime around Sgr A* is the ZV spacetime, including the Schwarzschild spacetime if we neglect the effects of the rotation, the surrounding matter, or the inclination of the orbital plane.}
  \label{fig:plot_f}
\end{figure}

We plot $f_\text{SP}$~\eqref{func_f_gamma} for the S2-Sgr A* system in Fig.~\ref{fig:plot_f}. Fig.~\ref{fig:plot_f_overall} shows an overall plot of $f_\text{SP}$. The region $f_\text{SP} < 0.91$, which is excluded by the observation, is filled in grey. Fig.~\ref{fig:plot_f_nmv} shows an enlarged plot of Fig.~\ref{fig:plot_f_overall} near the maximum value, and Fig.~\ref{fig:plot_f_gamma} is the plot with $1/\gamma$ as the horizontal axis. Table~\ref{table:paras_consts} in Appendix~\ref{app:paras} presents the orbital parameters of S2-Sgr A* that we have used to plot. 

From Fig.~\ref{fig:plot_f}, we immediately see that $f_\text{SP}$ has a maximum value within the range allowed by observations (i.e., $0.91 \lesssim f_\text{SP} \lesssim 1.29$). From Eq.~\eqref{f_max}, the maximum value can be analytically calculated as $f_\text{SP,max} \simeq 1.00091$ for $\gamma \rightarrow \infty$ ($\tilde{M}_2 \rightarrow -1/3$), which corresponds to the CC limit. Therefore, if future improvements in observational technology more strongly constrain the value of $f_\text{SP}$, and if $f_\text{SP} \gtrsim 1.00091$ is found to be within error, the possibility that the spacetime around Sgr A* is the ZV spacetime can be ruled out. For example, if the best fit value remains at $f_\text{SP} \simeq 1.1$ and the error in the observation becomes less than half of the current value, we can rule out the ZV assumption. Furthermore, the observational constraint $f_\text{SP} \gtrsim 0.91$ allows us to constrain a lower bound on the value of $\gamma$ as $\gamma \gtrsim 1.9 \times 10^{-2}$, which corresponds to $\tilde{M}_2 \lesssim 9.7 \times 10^2$ (see Figs.~\ref{fig:plot_f_overall} and~\ref{fig:plot_f_gamma}), whereas $-1/3 \leq \tilde{M}_2$ is required since the ZV solution is assumed. We summarize the characteristic values of $\gamma$, $\tilde{M}_2$, and $f_\text{SP}$ in Table~\ref{table:f}. 

\begin{table}[htbp]
\caption{The characteristic values of $\gamma$, $\tilde{M}_2$, and $f_\text{SP}$ for S2-Sgr A*. The $f_\text{SP}$ has maximum value $f_\text{SP,max} \simeq 1.00091$. If future observations find that $f_\text{SP} > f_\text{SP,max} \simeq 1.00091$, the possibility that the spacetime around Sgr A* is the ZV spacetime will be ruled out. Also, from the observational constraint $f_\text{SP} \gtrsim 0.91$, the lower limit of $\gamma$ is constraind to $\gamma \gtrsim 1.9 \times 10^{-2}$, which corresponds to $\tilde{M}_2 \lesssim 9.7 \times 10^2$.}
\centering
\begin{tabular}{@{\hspace{2truemm}} c @{\hspace{4truemm}} c @{\hspace{4truemm}} c @{\hspace{2truemm}} | @{\hspace{2truemm}}  c @{\hspace{2truemm}}}  \hline \hline
   $\gamma$ & $\tilde{M}_2$ & $f_\text{SP}$ & Remarks \\ \hline
   $1.9 \times 10^{-2}$ & $9.7 \times 10^2$ & 0.91 & Observational lower limit of $f_\text{SP}$ \\
   1 & 0 & 1.00088 & Schwarzschild case \\
   $\infty$ & $-1/3$ & 1.00091 & Theoretical upper limit of $f_\text{SP}$ (CC limit) \\
   \hline \hline
 \end{tabular}
 \label{table:f}
\end{table}

We should note that the above analysis is based on some assumptions, such as static, vacuum, and equatorial conditions. For a more realistic analysis, it would be necessary to take into account contributions from the rotation of the source, surrounding matter distributions, or the orbital inclination.

\section{New series representation for the periapsis shift} \label{sec:another_series}
In the future, with improvements in observational technology, we may observe the periapsis shift of stars in strong gravitational fields, where the PN expansion formula is not sufficiently accurate. Motivated by this, in this section, we derive a new series representation for the periapsis shift in the ZV spacetime using the recently proposed prescription~\cite{Katsumata:2024qzv}.
\subsection{Derivation}
The following derivation follows the prescription proposed in~\cite{Katsumata:2024qzv}. First, expanding $F(\chi)$~\eqref{eq:F_chi} in powers of the eccentricity $\bar{e}$, we obtain
\begin{align}
    F(\chi) = \sqrt{ \frac{ \left[ 1-M/(\gamma \bar{p}) \right]^{2(1-\gamma^2)} \left[ 1-2M/(\gamma \bar{p}) \right]^{\gamma^2}}{1 - 2(3 + 1/\gamma) M/\bar{p} + 2 (1 + 1/\gamma)(2 + 1/\gamma) (M/\bar{p})^2} } + \mathcal{O}(\bar{e}).
\end{align}
Factoring out the zeroth order term in this expression, Eq.~\eqref{peri_seki2} is rewritten as
\begin{align}
    \delta \phi_\text{ZV} = 2 \sqrt{ \frac{\left[ 1-M/(\gamma \bar{p}) \right]^{2(1-\gamma^2)} \left[ 1-2M/(\gamma \bar{p}) \right]^{\gamma^2}}{1 - 2(3 + 1/\gamma) M/\bar{p} + 2 (1 + 1/\gamma)(2 + 1/\gamma) (M/\bar{p})^2} } \int_{-\pi/2}^{\pi/2} G(\chi) \dd{\chi}, \label{peri_seki3}
\end{align}
where
\begin{align}
    G(\chi) := \sqrt{ \frac{1 - 2(3 + 1/\gamma) M/\bar{p} + 2 (1 + 1/\gamma)(2 + 1/\gamma) (M/\bar{p})^2}{\left[ 1-M/(\gamma \bar{p}) \right]^{2(1-\gamma^2)} \left[ 1-2M/(\gamma \bar{p}) \right]^{\gamma^2}} } F(\chi).
\end{align}
Then, expanding the integrand $G(\chi)$ in powers of $\bar{e}$, we can obtain
\begin{align}
    G(\chi) = \sum_{n=0}^{N-1} Y_n(\chi) \varepsilon^n \quad \text{(} N=1,2,3,\ldots \ \ \text{and} \ \ n=0,1,2,\ldots \text{)}, \label{G_expand}
\end{align}
where we have defined the new expansion parameter as
\begin{align}
    \varepsilon := \frac{\bar{e} M/\bar{p}}{\left[1-M/(\gamma \bar{p})\right] \left[1-2M/(\gamma \bar{p})\right] \left[ 1 - 2 (3+1/\gamma) M/\bar{p} + 2 \left(1+1/\gamma \right) \left(2+1/\gamma \right) (M/\bar{p})^2 \right] }. \label{def_epsilon}
\end{align}
Note that this expansion parameter $\varepsilon$ recovers that in the CC case (see Eq.~(3.28) in~\cite{Katsumata:2024qzv}) in the limit of $\gamma \rightarrow \infty$. The concrete forms of the coefficients $Y_n(\chi)$ are omitted here for their length and presented in Appendix~\ref{Yn_forms}. Then, from Eqs.~\eqref{peri_seki3} and~\eqref{G_expand}, we obtain
\begin{align}
    \delta \phi_\text{ZV} = 2 \sqrt{ \frac{ \left[ 1-M/(\gamma \bar{p}) \right]^{2(1-\gamma^2)} \left[ 1-2M/(\gamma \bar{p}) \right]^{\gamma^2}}{1 - 2(3 + 1/\gamma) M/\bar{p} + 2 (1 + 1/\gamma)(2 + 1/\gamma) (M/\bar{p})^2} } \sum_{n=0}^{N-1} \varepsilon^n \int_{-\pi/2}^{\pi/2} Y_n (\chi) \dd{\chi}. \label{peri_seki4}
\end{align}

Finally, let us evaluate the integral in Eq.~\eqref{peri_seki4} and obtain a new series representation for the periapsis shift. We can confirm that $Y_n$ with odd $n$ up to at least the seventh order (i.e., $Y_1$, $Y_3$, $Y_5$, and $Y_7$) are odd functions in terms of $\chi$, and therefore these coefficients do not contribute to the integral. Then, Eq.~\eqref{peri_seki4} can be integrated as
\begin{align}
    \delta \phi_\text{ZV} = 2 \pi \sqrt{ \frac{ \left[ 1-M/(\gamma \bar{p}) \right]^{2(1-\gamma^2)} \left[ 1-2M/(\gamma \bar{p}) \right]^{\gamma^2}}{1 - 2(3 + 1/\gamma) M/\bar{p} + 2 (1 + 1/\gamma)(2 + 1/\gamma) (M/\bar{p})^2} } \sum_{n=0}^{N-1} Z_{2n} \varepsilon^{2n}, 
\end{align}
where
\begin{align}
    Z_0 &:= 1, \\[8pt]
    Z_2 &:= \frac{1}{12} \Bigl[ -9 (3-4/\gamma) +6 (41 -19/\gamma -35/\gamma^2 -2/\gamma^3 ) M/\bar{p} \notag \\
    &\quad -3 (163 +218/\gamma -290/\gamma^2 -194/\gamma^3 -20/\gamma^4) (M/\bar{p})^2 \notag \\
    &\quad +6 \left( 74 +206/\gamma +37/\gamma^2 -276/\gamma^3 -153/\gamma^4 -20/\gamma^5 \right) (M/\bar{p})^3 \notag \\
    &\quad -2 (1+ 1/\gamma) (86 +292/\gamma +90/\gamma^2 -345/\gamma^3 -299/\gamma^4 -58/\gamma^5 ) (M/\bar{p})^4 \notag \\
    &\quad -8 (1+1/\gamma) (2+1/\gamma) ( 3 - 14/\gamma + 6/\gamma^2 + 6/\gamma^3 + 6/\gamma^4 + 5/\gamma^5) (M/\bar{p})^5 \notag \\
    &\quad + 4 (1+1/\gamma)^2 (2+1/\gamma)^2 ( 3 - 6/\gamma + 8/\gamma^2 - 3 /\gamma^3 - 1/\gamma^4) (M/\bar{p})^6 \Bigr],
\end{align}
and so forth. The concrete forms of the higher order coefficients, $Z_4, Z_6,\ldots$, are omitted here because of their length. In principle, it is also possible to obtain any higher order coefficients. 

Then, we finally obtain a new series representation for the periapsis shift per round in the ZV spacetime as
\begin{align}
    \Delta \phi_\text{ZV}(N) = 2 \pi \left\{ \sqrt{ \frac{\left[ 1-M/(\gamma \bar{p}) \right]^{2(1-\gamma^2)} \left[ 1-2M/(\gamma \bar{p}) \right]^{\gamma^2}}{1 - 2(3 + 1/\gamma) M/\bar{p} + 2 (1 + 1/\gamma)(2 + 1/\gamma) (M/\bar{p})^2} } \sum_{n=0}^{N-1} Z_{2n} \varepsilon^{2n} - 1 \right\}. \label{eq:ZV_peri_nr}
\end{align}
Note that this formula reproduces that for the Schwarzschild spacetime (see Eq.~(4) in~\cite{walters2018simple} or Eq.~(A.23) in~\cite{Katsumata:2024qzv}) for $\gamma=1$ and that for the CC spacetime (see Eq.~(3.33) in~\cite{Katsumata:2024qzv}) for $\gamma \rightarrow \infty$. As noted below Eqs.~\eqref{ER_e_d_p} and~\eqref{eq:PN_CC}, this is because the Erez-Rosen radial coordinate coincides with the Schwarzschild and Weyl radial coordinates for $\gamma=1$ and $\gamma \rightarrow \infty$, respectively (see also Appendix~\ref{app:review_ZV}), and therefore the definitions of the eccentricity and the semi-latus rectum coincide. Expanding Eq.~\eqref{eq:ZV_peri_nr} in powers of $M/\bar{p}$, we can also reproduce the PN expansion formula~\eqref{ZV_peri_PN}. Furthermore, in the case of $\bar{e}=0$, Eq.~\eqref{eq:ZV_peri_nr} recovers the exact formula for the periapsis shift of a quasi-circular orbit~\eqref{peri_qc}.

\subsection{Behavior of the expansion parameter}  \label{subsec:behavior_para}
\begin{figure}[htbp]
\hspace{-6truemm}
  \begin{minipage}{0.45\linewidth}
    \centering
    \includegraphics[keepaspectratio, scale=0.48]{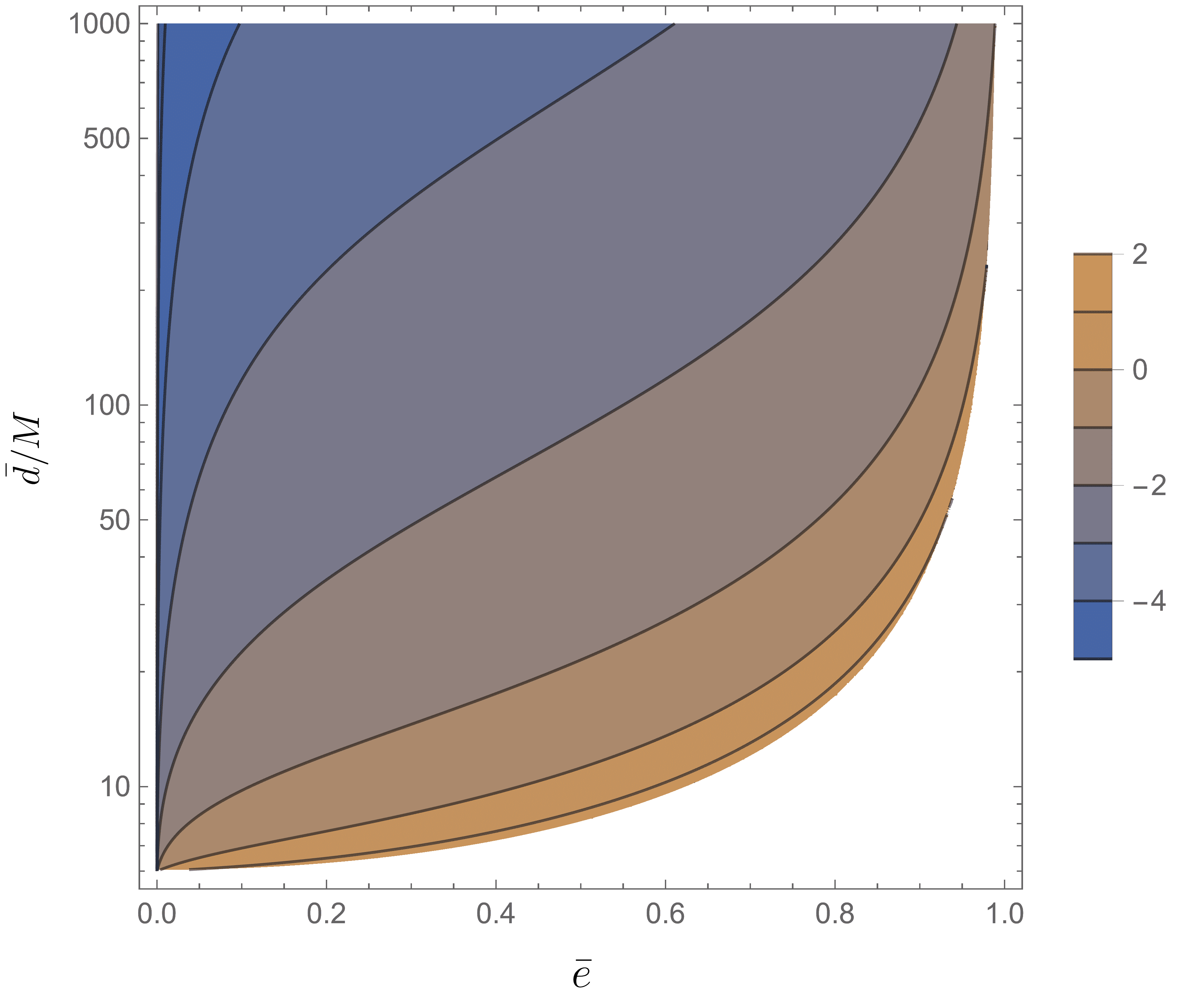}
    \subcaption{New expansion parameter $\varepsilon$}
    \label{fig:para_contour_a}
  \end{minipage} 
  \hspace{13truemm}
  \begin{minipage}{0.45\linewidth}
    \centering
    \includegraphics[keepaspectratio, scale=0.48]{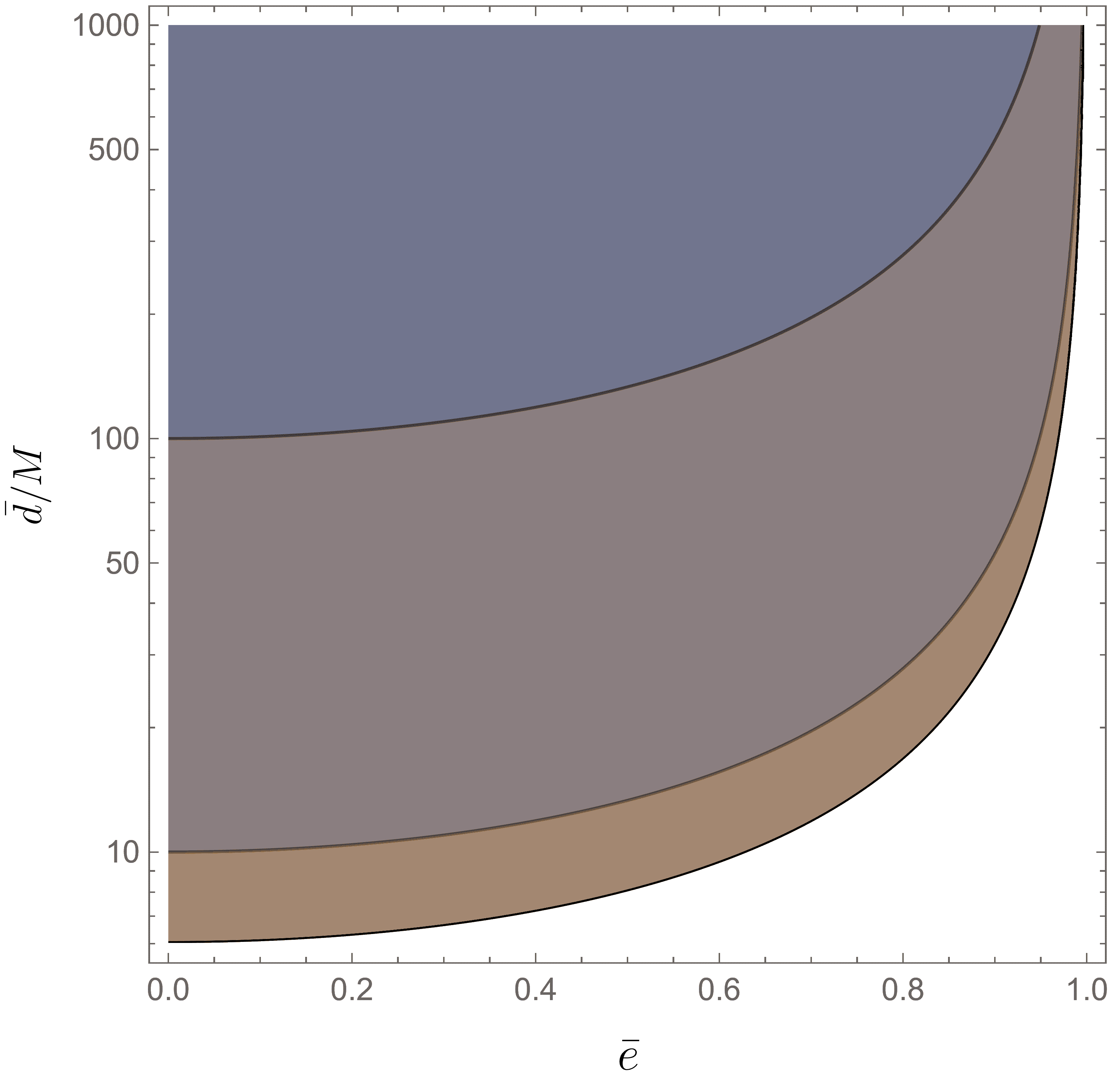}
    \subcaption{PN expansion parameter $M/\bar{p}$}
    \label{fig:para_contour_b}
  \end{minipage}
  \caption{The contour plots of the new expansion parameter $\varepsilon$ and the PN expansion parameter $M/\bar{p}=M/[\bar{d}(1-\bar{e}^2)]$. The base 10 logarithms of the parameters are plotted. In the case of $\bar{e} \to 0$, $\varepsilon$ approaches zero, while the PN expansion parameter approaches a non-zero value. It can be seen that the larger $\bar{d}/M$ also leads to the smaller both parameters. The deformation parameter $\gamma$ is fixed to be $\gamma=1/2$ as a non-trivial example. Note that the parameter regions that do not satisfy the conditions $1 - 2 (3+1/\gamma) M/\bar{p} + 2 (1+1/\gamma) (2+1/\gamma ) (M/\bar{p})^2 >0$, $\tilde{E}^2 \geq 0$, and $\tilde{L}^2 \geq 0$ are filled with white.}
  \label{fig:para_contour}
\end{figure}

Here, let us discuss the dependence of the new expansion parameter $\varepsilon$ on the orbital parameters. We show the contour plots of $\varepsilon$ and the PN expansion parameter $M/\bar{p}=M/[\bar{d}(1-\bar{e}^2)]$ in Fig.~\ref{fig:para_contour}. One can see that both parameters decrease as $\bar{e}$ is decreased. It is noteworthy that the major difference in the $\bar{e}$-dependence appears at $\bar{e} \rightarrow 0$. In this case, although the PN expansion parameter approaches a non-zero value $M/\bar{d}$, $\varepsilon$ approaches zero. It can be seen that the larger $\bar{d}/M$ also leads to the smaller $\varepsilon$ and the PN expansion parameter. These properties are similar to those for the Kerr and the CC cases in \cite{Katsumata:2024qzv}.

\begin{figure}[hbtp]
  \centering
  \includegraphics[scale=1.1]{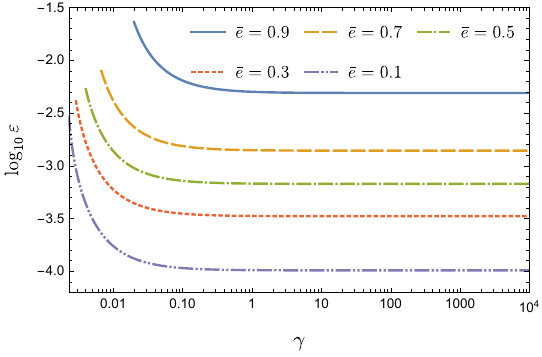}
  \caption{The $\gamma$-dependence of the new expansion parameter $\varepsilon$. The semi-major axis is fixed to be $\bar{d}/M=1000$. The base 10 logarithm of $\varepsilon$ is plotted. In the region $\gamma \gtrsim 0.1$, the value of $\varepsilon$ changes little compared to the region $\gamma \lesssim 0.1$. On the other hand, in the region $\gamma \lesssim 0.1$, we can see that $\varepsilon$ increases rapidly as $\gamma$ decreases. The lower end of $\gamma$ is determined by the conditions $\tilde{E}^2 \geq 0$ and $\tilde{L}^2 \geq 0$.}
  \label{fig:para_gamma_dep}
\end{figure}

The $\gamma$-dependence of the new expansion parameter $\varepsilon$ is also presented in Fig.~\ref{fig:para_gamma_dep}. We can see that $\varepsilon$ monotonically decreases with increasing $\gamma$. In the region $\gamma \gtrsim 0.1$, the value of $\varepsilon$ changes less compared to the region $\gamma \lesssim 0.1$. On the other hand, in the region $\gamma \lesssim 0.1$, we can see that $\varepsilon$ increases rapidly as $\gamma$ decreases. We can understand this property by the analytical expression of $\varepsilon$~\eqref{def_epsilon}: As $\gamma$ becomes smaller, the part of the denominator, $1-2M/(\gamma \bar{p})$, approaches zero most quickly, and $\varepsilon$ diverges. As we will see in the next subsection, this divergence of the expansion parameter $\varepsilon$ reduces the accuracy of the new series expansion for small $\gamma$.

\subsection{Accuracy of the truncated new formula}
Let us focus on the dependence of the accuracy of the formula that is obtained by truncating the new series representation~\eqref{eq:ZV_peri_nr} up to a finite number of terms, and compare it with that of the truncated PN expansion formula. Now, we define the periapsis shift per round as the integral form:
\begin{align}
    I := 2 \int_{-\pi/2}^{\pi/2} F(\chi) \dd{\chi} - 2 \pi.
\end{align}
In addition, let us summarize the PN expansion formula~\eqref{ZV_peri_PN} as:
\begin{align}
    \Delta \phi_\text{ZV,PN}(N) := \sum_{n=1}^N \bar{\Delta} \phi_{\text{ZV},n \text{PN}} \quad (N=1,2,3,\ldots).
\end{align}
The explicit formula up to $N=6$ is shown as Eq.~\eqref{ZV_peri_PN_high} in Appendix~\ref{ZV_PN_high_order}. Furthermore, we define the relative errors of the truncated new formula and the truncated PN expansion formula as
\begin{align}
  \Delta_\text{ZV,NF}(N) := \biggl| \frac{\Delta \phi_\text{ZV} (N)}{I} -1 \biggr| , \quad 
  \Delta_\text{ZV,PN} (N) := \biggl| \frac{\Delta \phi_\text{ZV,PN} (N) }{I} -1 \biggr|,
\end{align}
respectively.

\begin{figure}[htbp]
\hspace{-10truemm}
  \begin{minipage}{0.45\linewidth}
    \centering
    \includegraphics[keepaspectratio, scale=1.09]{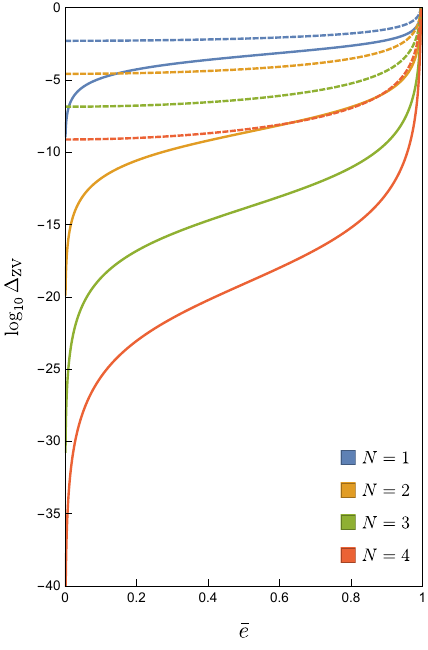}
    \subcaption{$\bar{d}/M=1000$}
    \label{fig:gosaeplotd1000}
  \end{minipage} 
  \hspace{7truemm}
  \begin{minipage}{0.45\linewidth}
    \centering
    \includegraphics[keepaspectratio, scale=1.09]{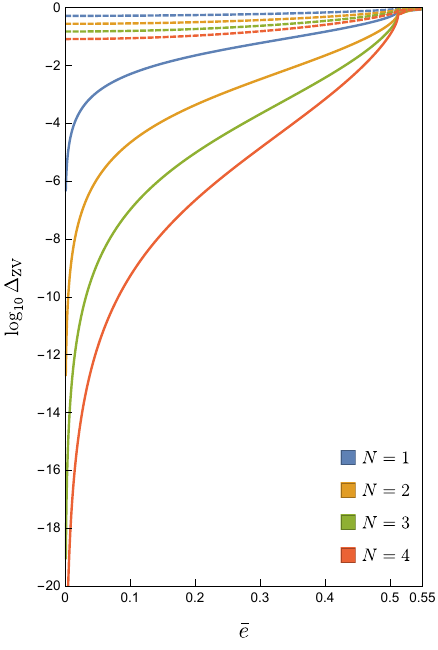}
    \subcaption{$\bar{d}/M=10$}
    \label{fig:gosaeplotd10}
  \end{minipage}
  \caption{The dependence of the relative errors of the truncated new formula $\Delta_\text{ZV,NF}(N)$ and the truncated PN expansion formula $\Delta_\text{ZV,PN}(N)$ on the eccentricity $\bar{e}$. The solid and dashed lines represent the errors $\Delta_\text{ZV,NF}(N)$ and $\Delta_\text{ZV,PN}(N)$, respectively. The parameter $\gamma$ is fixed to be $\gamma=1/2$ as an example. It can be seen that in the case of $\bar{e} \to 0$, the errors of the truncated new formula approach zero, while the errors of the truncated PN expansion formula approach non-zero values.}
  \label{fig:gosaeplot}
\end{figure}

\begin{figure}[htbp]
\hspace{-10truemm}
  \begin{minipage}{0.45\linewidth}
    \centering
    \includegraphics[keepaspectratio, scale=1.09]{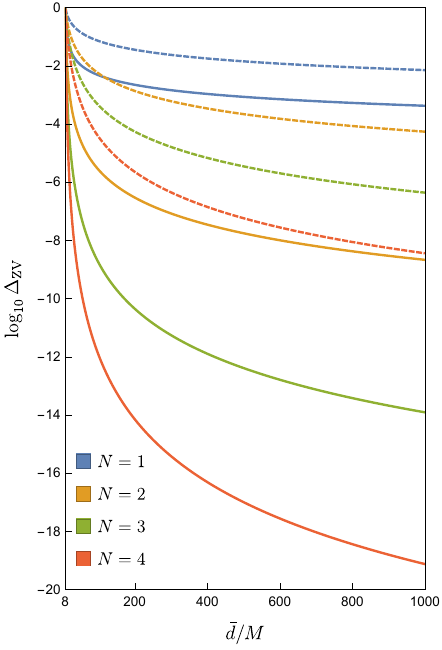}
    \subcaption{$\bar{e}=1/2$}
    \label{fig:gosadplote05}
  \end{minipage} 
  \hspace{7truemm}
  \begin{minipage}{0.45\linewidth}
    \centering
    \includegraphics[keepaspectratio, scale=1.09]{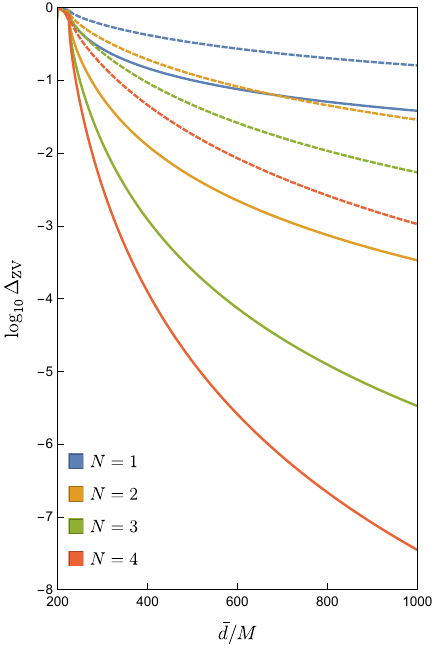}
    \subcaption{$\bar{e}=0.98$}
    \label{fig:gosadplote098}
  \end{minipage}
  \caption{The dependence of the relative errors of the truncated new formula $\Delta_\text{ZV,NF}(N)$ and the truncated PN expansion formula $\Delta_\text{ZV,PN}(N)$ on the semi-major axis $\bar{d}$. The solid and dashed lines represent the errors $\Delta_\text{ZV,NF}(N)$ and $\Delta_\text{ZV,PN}(N)$, respectively. The parameter $\gamma$ is fixed to be $\gamma=1/2$ as an example. It can be seen that the error of the truncated new formula decreases as $\bar{d}/M$ is increased.}
  \label{fig:gosadplot}
\end{figure}

\begin{figure}[hbtp]
  \centering
  \includegraphics[scale=1.3]{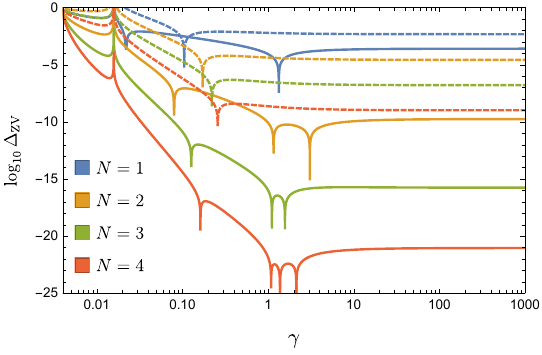}
  \caption{The dependence of the relative errors of the truncated new formula $\Delta_\text{ZV,NF}(N)$ and the truncated PN expansion formula $\Delta_\text{ZV,PN}(N)$ on $\gamma$. The solid and dashed lines represent the errors $\Delta_\text{ZV,NF}(N)$ and $\Delta_\text{ZV,PN}(N)$, respectively. The semi-major axis and the eccentricity are fixed as $\bar{d}/M=1000$ and $\bar{e}=1/2$, respectively. For $\gamma \lesssim 1$, we can see that the error of the truncated new formula decreases as $\gamma$ is increased. The error is large if $\gamma$ is much smaller than unity. The lower limit $\gamma = 0.004$ is determined by the conditions $\tilde{E}^2 \geq 0$ and $\tilde{L}^2 \geq 0$.}  
  \label{fig:gosagammaplot}
\end{figure}

Figs.~\ref{fig:gosaeplot}-\ref{fig:gosagammaplot} show the dependence of the relative errors of the truncated new formula $\Delta_\text{ZV,NF}(N)$ and truncated PN expansion formula $\Delta_\text{ZV,PN}(N)$ on the eccentricity $\bar{e}$, the semi-major axis $\bar{d}$, and the deformation parameter $\gamma$, respectively. First, let us focus on the $\bar{e}$-dependence. Fig.~\ref{fig:gosaeplotd1000} shows that the error of the truncated new formula (solid lines) decreases monotonically as $\bar{e}$ decreases, with larger $N$ further improving accuracy. However, the accuracy declines rapidly as $\bar{e}$ approaches 1. Note that the error of the truncated PN expansion formula (dashed line) shows similar properties. We emphasize that the truncated new formula has better accuracy than the truncated PN expansion formula with smaller $N$. For example, it can be seen that the truncated new formula with $N=2$ has an accuracy comparable to that of the truncated PN expansion formula with $N=4$. A significant difference in the $\bar{e}$-dependence of the error between the truncated new formula and the truncated PN expansion formula appears at $\bar{e} \rightarrow 0$: the error of the truncated new formula rapidly approaches zero, while the error of the truncated PN expansion formula approaches non-zero values. This property is also seen in the Kerr and the CC cases~\cite{Katsumata:2024qzv}, and it would be due to the $\bar{e}$-dependence of the expansion parameter $\varepsilon$ discussed in Sec.~\ref{subsec:behavior_para}. In addition, Fig.~\ref{fig:gosaeplotd10} shows the errors for orbits closer to the center ($\bar{d}/M=10$). It can be seen that the error of the truncated PN expansion formula is at most about $10^{-1}$ even when $N=4$, whereas the error of the truncated new formula has higher accuracy than the PN expansion formula if $\bar{e}$ is sufficiently small. It is one of the advantages of the new series representation that it is more accurate than the PN expansion formula, even if the orbit is close to the center, where the convergence of the PN expansion is not guaranteed.

Next, we discuss the $\bar{d}$-dependence of the errors. Fig.~\ref{fig:gosadplote05} shows that the larger $\bar{d}/M$ and $N$ reduce the error of the truncated new formula (solid lines) more effectively than that of the truncated PN expansion formula (dashed lines). Furthermore, Fig.~\ref{fig:gosadplote098} shows the errors for a highly eccentric orbit ($\bar{e}=0.98$). One can see that the truncated new formula also achieves higher accuracy if $\bar{d}/M$ and $N$ are sufficiently large. One of the advantages of the truncated new formula is that even if the orbit is highly eccentric, it has higher accuracy than the truncated PN expansion formula if $\bar{d}/M$ and $N$ are sufficiently large.

Finally, we discuss the $\gamma$-dependence of the errors. From Fig.~\ref{fig:gosagammaplot}, for $\gamma \lesssim 1$, it can be found that the error of the truncated new formula decreases as $\gamma$ increases (solid lines). On the other hand, for $\gamma \gtrsim 1$, the error does not change much more than the region $\gamma \lesssim 1$. This $\gamma$-dependence of the error of the truncated new formula is supposed to reflect the $\gamma$-dependence of the expansion parameter $\varepsilon$ discussed in Sec.~\ref{subsec:behavior_para} (see also Fig.~\ref{fig:para_gamma_dep}). We should note that, in the region where $\gamma$ is very small ($\gamma \sim 0.01$), the error seems to diverge and the accuracy of the new series expansion deteriorates. This would be due to the divergence of the new expansion parameter $\varepsilon$ for small $\gamma$ discussed in Sec.~\ref{subsec:behavior_para}. We also mention that, as $\gamma$ sufficiently grows, the error values converge to the results in the CC case discussed in~\cite{Katsumata:2024qzv} (see Fig.~8a in~\cite{Katsumata:2024qzv}).

\section{Summary and conclusions}
\label{sec:sum_conc}
In this paper, we have discussed the periapsis shift in the equatorial plane in the ZV spacetime. We have derived the PN expansion formula for the periapsis shift for arbitrary eccentricity in this spacetime~\eqref{ZV_peri_PN}. The contribution of the deformation parameter $\gamma$ appears from the 2PN order term. The obtained PN expansion formula recovers those in the Schwarzschild spacetime for $\gamma=1$ and the CC spacetime for $\gamma \rightarrow \infty$. We have investigated the deviation of the obtained PN expansion formula from that in the Schwarzschild spacetime. The deviation $\delta_\text{ZV-S}$~\eqref{devi_ZV_S} is the linear function of $1/\gamma^2$ or $\tilde{M}_2$ and has the maximum value for the CC limit.

Furthermore, we have discussed the constraining of the deformation parameter $\gamma$ and the corresponding quadrupole moment $\tilde{M}_2$ of Sgr A* using the obtained PN expansion formula and the observational data on the periapsis shift of S2~\cite{GRAVITY:2020gka}. We have found that $f_\text{SP}$ has the maximum value $f_\text{SP,max} \simeq 1.00091$ for $\gamma \rightarrow \infty$ ($\tilde{M}_2 \rightarrow -1/3$). Therefore, if future observations find $f_\text{SP} \gtrsim 1.00091$ within error, the possibility that the spacetime around Sgr A* is the ZV spacetime can be excluded. For example, if the best fit value remains at $f_\text{SP} \simeq 1.1$ and the error in the observation becomes less than half of the current value, the exclusion can be achieved. Furthermore, from the observational constraint $f_\text{SP} \gtrsim 0.91$, we have obtained the lower bound of $\gamma$ as $\gamma \gtrsim 1.9 \times 10^{-2}$, which is interpreted as $\tilde{M}_2 \lesssim 9.7 \times 10^2$. We should note that, although the present analysis considers the equatorial orbits in the ZV spacetime (static and vacuum), the effects of the rotation of the source, matter distribution, or the inclination of the orbital plane should also contribute to the periapsis shift. Therefore, for a more realistic analysis, the discussion should be conducted using stationary spacetimes such as the Tomimatsu-Sato or the $\delta$-Kerr spacetimes.  It would also be interesting to consider the effects of matter distributions or to remove the assumption of being on the equatorial plane. These are our future works.

Finally, we have derived the new series representation for the periapsis shift in the ZV spacetime using the recently proposed prescription~\cite{Katsumata:2024qzv}. According to the prescription, the expansion parameter $\varepsilon$ is defined as the eccentricity divided by the dimensionless quantity that vanishes in the ISCO limit. This means that the parameter $\varepsilon$ denotes both the eccentricity of the orbit and its proximity to the ISCO. We have examined the behaviors of the expansion parameter $\varepsilon$ and the relative error of the truncated series representation. The parameter $\varepsilon$ decreases as the eccentricity $\bar{e}$ is decreased. At $\bar{e} \rightarrow 0$, although the PN expansion parameter approaches a non-zero value $M/\bar{d}$, $\varepsilon$ approaches zero. It can also be seen that the larger $\bar{d}/M$ and $\gamma$ also lead to the smaller $\varepsilon$. The relative error of the truncated new formula also has a similar dependence to that of $\varepsilon$: the relative error of the truncated new formula approaches zero in the case of $\bar{e} \rightarrow 0$, while the error of the truncated PN expansion formula approaches a non-zero value. The larger $\bar{d}/M$ also leads to a smaller error of the truncated new formula. Also, for $\gamma \lesssim 1$, the error decreases as $\gamma$ increases.

Qualitatively, the smaller the eccentricity, the higher the accuracy of the truncated new formula. If the eccentricity is sufficiently small, the truncated new formula has higher accuracy than the truncated PN expansion formula, even in the strong-field regime where the convergence of the PN expansion formula is not guaranteed. In addition, even if the orbit is highly eccentric, the truncated new formula has comparable or higher accuracy than the truncated PN expansion formula if the semi-major axis is sufficiently large. The simple and highly accurate formula obtained by truncating the new series representation is practical and useful, for example, for discussing the periapsis shifts in the strong-field regime and for verifying numerical calculations. Further applications of the new series expansion for the periapsis shift to other spacetimes are to be investigated.

\acknowledgments
The authors are grateful to L. Fern\'{a}ndez-Jambrina, A. Idrissov, T. Igata, H. Iizuka, S. Ishikawa, M. Kimura, T. Kobayashi, H. Maeda, B. Mirza, K. Nakao, K. Ogasawara, H. Saida, S. Vagnozzi, and H. Yoshino for their helpful comments. This work was supported by Rikkyo University Special Fund for Research (A.K.), JSPS KAKENHI Grants Nos.~JP20H05853 and JP24K07027 (T.H.).
\renewcommand{\theequation}{A.\arabic{equation}}
\makeatletter
\@addtoreset{equation}{section}
\makeatother
\appendix
\section{PN expansion formula with higher order terms} \label{ZV_PN_high_order}
In this appendix, we present the PN expansion formula for the periapsis shift in the ZV spacetime with higher PN order terms:
\begin{align}
    \Delta \phi_\text{ZV} &= \frac{6 \pi M}{\bar{p}} + \left[ \left(44+\frac{12}{\gamma}-\frac{2}{\gamma^2}\right)+3\left(-3+\frac{4}{\gamma}\right) \bar{e}^2 \right] \frac{\pi M^2}{2 \bar{p}^2} \notag \\[5pt]
    &\hspace{-10truemm} + \left[\left(192+\frac{88}{\gamma}-\frac{6}{\gamma^2}-\frac{4}{\gamma ^3}\right)+\left(-53+\frac{52}{\gamma}+\frac{50}{\gamma^2}-\frac{4}{\gamma^3}\right) \bar{e}^2 \right] \frac{\pi M^3}{2 \bar{p}^3} \notag \\[5pt]
    &\hspace{-10truemm} + \biggl[\left(14064+\frac{9216}{\gamma}+\frac{144}{\gamma^2} -\frac{672}{\gamma^3}-\frac{72}{\gamma ^4}\right)+\left(-4096+\frac{4976}{\gamma}+\frac{5592}{\gamma^2}+\frac{1408}{\gamma^3}-\frac{320}{\gamma^4}\right) \bar{e}^2 \notag \\
    &\hspace{-10truemm} +\left(227-\frac{848}{\gamma}+\frac{342}{\gamma^2}+\frac{416}{\gamma^3}-\frac{32}{\gamma^4}\right) \bar{e}^4 \biggr] \frac{\pi M^4}{32 \bar{p}^4} \notag \\[5pt]
    &\hspace{-10truemm} + \biggl[ \left(992880+\frac{843840}{\gamma}+\frac{94320}{\gamma^2}-\frac{75840}{\gamma^3}-\frac{17640}{\gamma^4} -\frac{480}{\gamma^5} \right) \notag \\
    &\hspace{-10truemm} + \left(-250320+\frac{475200}{\gamma}+\frac{627240}{\gamma^2}+\frac{175920}{\gamma^3}+\frac{6480}{\gamma^4}-\frac{13920}{\gamma^5}\right) \bar{e}^2 \notag \\
    &\hspace{-10truemm} + \left(11651-\frac{95640}{\gamma}+\frac{5330}{\gamma^2}+\frac{91680}{\gamma^3}+\frac{35264}{\gamma^4}-\frac{5760}{\gamma^5}\right) \bar{e}^4
    \biggr] \frac{\pi M^5}{480 \bar{p}^5} \notag \\[5pt]
    &\hspace{-10truemm} + \biggl[ \left( 57138240+\frac{59572800}{\gamma}+\frac{12359520}{\gamma^2}-\frac{5400000}{\gamma^3}-\frac{2274240}{\gamma^4}-\frac{159840}{\gamma^5} +\frac{10800}{\gamma^6} \right) \notag \\
    &\hspace{-10truemm} + \left(-8928240+\frac{38545920}{\gamma}+\frac{53192400}{\gamma^2}+\frac{19265760}{\gamma^3}+\frac{74760}{\gamma^4}-\frac{744480}{\gamma^5}-\frac{366720}{\gamma^6}\right) \bar{e}^2 \notag \\
    &\hspace{-10truemm} + \left(-1589380-\frac{7753212}{\gamma}+\frac{2236146}{\gamma^2}+\frac{9435960}{\gamma^3}+\frac{5409540}{\gamma^4}+\frac{1069632}{\gamma^5}-\frac{388736}{\gamma^6}\right) \bar{e}^4 \notag \\
    &\hspace{-10truemm} + \left(87645+\frac{139812}{\gamma}-\frac{538410}{\gamma^2}-\frac{126120}{\gamma^3}+\frac{365400}{\gamma^4}+\frac{146688}{\gamma^5}-\frac{23040}{\gamma^6}\right) \bar{e}^6 \biggr] \frac{\pi M^6}{5760 \bar{p}^6}. \label{ZV_peri_PN_high}
\end{align}
The 1PN order term does not depend on $\gamma$, and coincides with those in the Schwarzschild, the CC, and the Kerr spacetimes, etc. The contribution of $\gamma$ appears from the 2PN order term. We should note that this formula is expressed by the eccentricity $\bar{e}$ and the semi-latus rectum $\bar{p}$, which are defined in terms of the Erez-Rosen radial coordinate. It is also important to note that Eq.~\eqref{ZV_peri_PN_high} can reproduce the formulae in the Schwarzschild spacetime (see e.g.~\cite{He:2023joa,Katsumata:2024qzv}) for $\gamma=1$ and in the CC spacetime~\cite{Bini:2005dy,Katsumata:2024qzv} for $\gamma \rightarrow \infty$. As mentioned below Eqs.~\eqref{ER_e_d_p},~\eqref{eq:PN_CC}, and~\eqref{eq:ZV_peri_nr}, this is because the Erez-Rosen radial coordinate coincide with the Schwarzschild radial coordinate for $\gamma = 1$ and with the Weyl radial coordinate for $\gamma \rightarrow \infty$ (see also Appendix~\ref{app:review_ZV}), which means that the definitions of the eccentricity and the semi-latus rectum coincide.

\renewcommand{\theequation}{B.\arabic{equation}}
\makeatletter
\@addtoreset{equation}{section}
\makeatother
\section{Orbital parameters of S2-Sgr A*} \label{app:paras}
Table~\ref{table:paras_consts} presents the orbital parameters of S2-Sgr A* we have used in Sec.~\ref{sec:constraint}. The values of the mass of Sgr A*, $M$, the semi-major axis, $a$, the eccentricity, $e$, and the distance to the galactic center, $R_0$, are taken from \cite{GRAVITY:2020gka}. The values of the heliocentric gravitational constant, $G M_\odot = 1.32712440041 \times 10^{20} \, \text{m}^3 \text{s}^{-2}$, the speed of light, $c = 299792458 \, \text{m} \, \text{s}^{-1}$, and the astronomical unit, $1.495978707 \times 10^{11} \, \text{m}$, are taken from \cite{rikanenpyou2025}. We have approximately calculated the semi-major axis $d$ (in units of length) using $a$ and $R_0$ as follows:
\begin{align}
   d \approx R_0 \times \left( a \times 10^{-3} \times \frac{1}{3600} \times \frac{\pi}{180} \right) = 1.542826... \times 10^{14} \, \text{m}.
\end{align}

\begin{table}[htbp]
\caption{The orbital parameters of S2-Sgr A*. The values of $M$, $a$, $e$, and $R_0$ are taken from \cite{GRAVITY:2020gka}.}
\centering
\begin{tabular}{ @{\hspace{2truemm}} c @{\hspace{4truemm}} c @{\hspace{4truemm}} c @{\hspace{4truemm}} c @{\hspace{2truemm}} }  \hline \hline
   & Symbol & Value & Unit \\ \hline
   Mass & $M$ & 4.261 & $10^6 M_\odot$ \\
   Semi-major axis & $a$ & 125.058 & mas \\
   (ditto) & $d$ & 1.542826... & $10^{14} \, \text{m} $\\
   Eccentricity & $e$ & 0.884649 & - \\
   Distance to the galactic center & $R_0$ & 8246.7 & pc \\
   \hline \hline
 \end{tabular}
 \label{table:paras_consts}
\end{table}

\renewcommand{\theequation}{C.\arabic{equation}}
\makeatletter
\@addtoreset{equation}{section}
\makeatother
\section{Periapsis shift of a quasi-circular orbit} \label{app:ps_qc}
In Sec.~\ref{sec:peri_PN}, we have derived the approximate formula for the periapsis shift in ZV spacetime for arbitrary eccentricity by the PN expansion. In this appendix, we focus on quasi-circular orbits and derive an exact formula for the periapsis shift in the ZV spacetime without using the PN expansion. Although an approximate formula for the periapsis shift is derived in~\cite{Chakrabarty:2022fbd} under the approximation that the orbital eccentricity is sufficiently small, the method we will use evaluates the periapsis shift of a quasi-circular orbit using the orbital angular velocity and the radial frequency, and provides an exact and simple formula, which is totally inconsistent with the formula obtained in~\cite{Chakrabarty:2022fbd}. 

\subsection{Overview of a quasi-circular orbit and the periapsis shift} \label{subsec:qc}
Before going into specific calculations, we will briefly introduce the quasi-circular orbits and the periapsis shift in those orbits. A quasi-circular orbit, which is also called a nearly circular orbit, can be regarded as a circular orbit undergoing infinitesimally small radial oscillations while maintaining a constant orbital angular velocity. As mentioned in Sec.~\ref{subsec:devi}, the qualitative discussion of the periapsis shift, illustrated there with the schematic illustration, can also be applied to the quasi-circular orbits. In the case of a quasi-circular orbit, the periapsis shift is characterized by the magnitude relationship of the orbital angular frequency $\omega_\phi$ and the radial frequency $\omega_r$. In Newtonian gravity, the periapsis is not shifted since these frequencies coincide (i.e., $\omega_\phi = \omega_r$), and the orbit closes in one revolution. In contrast, general relativistic effects make these frequencies differ, leading to the periapsis shift. In the case where the relativistic effects make the orbital angular frequency greater than the radial frequency (i.e., $\omega_\phi > \omega_r$), the orbit completes a revolution before reaching the next periapsis, so the periapsis is shifted in the same direction as the orbital motion (forward shift). Conversely, in this case where the orbital angular frequency is less than the radial frequency (i.e., $\omega_\phi < \omega_r$), the orbit completes a revolution after reaching the next periapsis, so the periapsis is shifted in the opposite direction to the orbital motion (backward shift).

\subsection{Exact formula for the periapsis shift of a quasi-circular orbit}
Using the orbital angular and radial frequencies, $\omega_\phi$ and $\omega_r$, an exact formula for the periapsis shift of the quasi-circular orbit per round is given as follows~\cite{Wald:1984rg,Igata:2022nkt,Igata:2022rcm,Harada:2022uae,Igata:2025trj}:
\begin{align}
    \Delta \phi_\text{qc} &:= 2 \pi \left( \frac{\omega_\phi}{\omega_r} -1 \right). \label{eq:qc_peri_pn_kousiki}
\end{align}
This formula can also be derived from the definition of the periapsis shift~\eqref{peri_ZV_teigi} as follows: From Eq.~\eqref{peri_ZV_teigi}, we have
\begin{align}
    \Delta \phi &= 2 \int_{r_p}^{r_a} \dv{\phi}{r} \dd{r} - 2 \pi = 2 \pi \left( \frac{1}{\pi} \int_{\phi_p}^{\phi_a} \dd{\phi} -1 \right) = 2 \pi \left( \frac{1}{\pi} \int_{\tau_p}^{\tau_p + T_r/2} \dot{\phi} \dd{\tau} -1 \right). \label{eq:peri_qc_def_c}
\end{align}
The integral in the third quantity represents the rotation angle from the periapsis to the apoapsis, where $\phi_p$ and $\phi_a$ denote the angles at the periapsis and apoapsis, respectively. In the rightmost term, the rotation angle is rewritten as an integral with respect to the particle's proper time $\tau$, where $\tau_p$ and $T_r$ denote the proper time when the particle is at the periapsis and the orbital period from the periapsis to the next periapsis (i.e., period of the simple harmonic motion in the radial direction), respectively. Here, let us focus on the quasi-circular orbits. In this case, while the motion oscillates harmonically in the radial direction, the angular velocity $\dot{\phi}$ remains constant. Hence, since $\dot{\phi}$ can be factored out of the integral, Eq.~\eqref{eq:peri_qc_def_c} can be calculated as
\begin{align}
    \Delta \phi_\text{qc} &= 2 \pi \left( \frac{\dot{\phi} T_r}{2 \pi} - 1 \right) = 2 \pi \left( \frac{\omega_\phi}{\omega_r} - 1 \right),
\end{align}
where the orbital period $T_r$ has been rewritten as $T_r = 2\pi/\omega_r$ using the radial frequency $\omega_r$. The above shows that Eq.~\eqref{eq:qc_peri_pn_kousiki} can be derived from the definition~\eqref{peri_ZV_teigi}.

\subsection{Case of the Zipoy-Voorhees spacetime}
\subsubsection{Circular orbits of massive test particles}
Before calculating the periapsis shift of a quasi-circular orbit in the ZV spacetime, we first consider timelike circular orbits in this spacetime as a preliminary. Let us rewrite Eq.~\eqref{rdot} as
\begin{align}
    \frac{1}{2} \dot{r}^2 + V(r) = 0, \label{rdot2}
\end{align}
where we have defined the effective potential as
\begin{align}
    V(r) := - \frac{1}{2} \left( \frac{f}{g} \right)^{1-\gamma^2} \left[ \tilde{E}^2 - f^{\gamma} \left( 1 + f^{\gamma-1} \frac{\tilde{L}^2}{r^2} \right) \right]. \label{eq:eff_po_ZV}
\end{align}
In addition, the Euler-Lagrange equation of the radial motion reduces to
\begin{align}
    \ddot{r} + V^\prime (r) = 0, \label{rdot3}
\end{align}
where the prime denotes the derivative with respect to $r$. Now, we focus on circular orbits with radius $r = r_0$. Taking Eqs.~\eqref{rdot2} and \eqref{rdot3} into account, the conditions for the circular orbits $\dot{r} = 0$ and $\ddot{r} = 0$ imply $V(r_0) = 0$ and $V^\prime (r_0) = 0$, respectively. Solving these equations, we get formulae for the energy and the angular momentum of the particle in the circular orbits as
\begin{align}
    \tilde{E}^2 = \frac{ \left[ 1 -  (1+ 1/\gamma) M/r_0 \right] \left[ 1-2M/(\gamma r_0) \right]^\gamma}{ 1 -  (2+1/\gamma) M/r_0}, \quad
    \tilde{L}^2 = \frac{M r_0 \left[ 1-2M/(\gamma r_0) \right]^{1-\gamma}}{1-(2+1/\gamma) M/r_0}, \label{circular_E_L}
\end{align}
respectively. Since $\tilde{E}^2 \geq 0$ and $\tilde{L}^2 \geq 0$, we can find that the conditions
\begin{align}
    r_0 &> \left(2+\frac{1}{\gamma}\right) M \quad \text{for} \ \ \gamma \geq \frac{1}{2}, \label{conditions_co1} \\[5pt]
    r_0 &> \frac{2M}{\gamma} \quad \text{for} \ \ 0 < \gamma \leq \frac{1}{2}, \label{conditions_co2}
\end{align}
must be satisfied for the existence of the circular orbit. Note that the surface $r=2M/\gamma$ corresponds to the curvature singularity~\cite{Kodama:2003ch}. For $\gamma \geq 1/2$, there exist photon circular orbits at $r=(2+1/\gamma)M$~\cite{Toshmatov:2019qih}, while for $0 < \gamma < 1/2$ there are no photon circular orbits.

Now, let us consider the situation where the circular orbit at $r = r_0$ is perturbed as $r = r_0 + \delta r$ with $\delta r$ being infinitesimally small. In this case, from Eq.~\eqref{rdot3}, we can obtain the equation of motion for $\delta r$, that is
\begin{align}
    \ddot{\delta r} + V^{\prime \prime}(r_0) \delta r = 0. \label{eq:qc_freq}
\end{align}
We can see that if $V^{\prime \prime}(r_0) \geq 0$, the circular orbit is stable, and it becomes a quasi-circular orbit, where the $\delta r$ obeys a simple harmonic motion. Taking the circular orbit conditions given by Eqs.~\eqref{conditions_co1} and~\eqref{conditions_co2} into account, the condition for the stable circular orbit, $V^{\prime \prime}(r_0) \geq 0$, implies
\begin{align}
    A(r_0) \geq 0, \label{eq:con_stable}
\end{align}
where we have defined the quadratic function $A(r_0)$ as
\begin{align}
    A(r_0) := r_0^2 - 2M \left(3+\frac{1}{\gamma} \right) r_0 + 2 M^2 \left( 1 +  \frac{1}{\gamma} \right) \left( 2 + \frac{1}{\gamma} \right).
\end{align}
Note that the sign of $A(r_0)$ and the sign of $V^{\prime \prime}(r_0)$ coincide. 

The discriminant $D$ of $A(r_0)$ can be calculated as
\begin{align}
    D = 4 \left( 5 - \frac{1}{\gamma^2} \right) M^2.
\end{align}
For $D<0$, i.e., $0< \gamma < 1/\sqrt{5}$, $A(r_0)$ is always positive, and therefore there exists a stable circular orbit for any $r_0 > 2M/\gamma$. On the other hand, for $D \geq 0$, i.e., $\gamma \geq 1/\sqrt{5} \simeq 0.447$, the quadratic equation, $A(r_0)=0$, which can be obtained from the limiting case of Eq.~\eqref{eq:con_stable}, has a real solution
\begin{align}
    r_\text{MSCO}^{(\pm)} = \left( 3+ \frac{1}{\gamma} \pm \sqrt{ 5 - \frac{1}{\gamma^2} } \right) M.
\end{align}
This denotes the radius of the marginally stable circular orbits in the ZV spacetime~\cite{Toshmatov:2019qih,Chowdhury:2011aa,Boshkayev:2015jaa,Abdikamalov:2019ztb}. Note that for $\gamma=1$ and $\gamma \rightarrow \infty$, $r_\text{MSCO}^{(+)}$ recovers the ISCO raii in the Schwarzschild spacetime, $r_\text{MSCO}^{(+)}=6M$, and in the CC spacetime~\cite{Gonzalez:2011fb}, $r_\text{MSCO}^{(+)} = (3+\sqrt{5})M$, respectively. We can find that $A(r_0) \geq 0$ for $r_0 \leq r_\text{MSCO}^{(-)}$ or $r_\text{MSCO}^{(+)} \leq r_0$, and therefore the condition for the stable circular orbits~\eqref{eq:con_stable} is satisfied in these regions. While no stable circular orbits can exist in $r_\text{MSCO}^{(-)} < r_0 < r_\text{MSCO}^{(+)}$. Furthermore, for $\gamma \geq 1/\sqrt{5}$, $r_\text{MSCO}^{(+)}$ always satisfies the circular orbit conditions \eqref{conditions_co1} and \eqref{conditions_co2}. One the other hand, $r_\text{MSCO}^{(-)}$ satisfies the circular orbit condition for $1/\sqrt{5} \leq \gamma \leq 1/2$, but not for $\gamma \geq 1/2$. This means that, for $1/\sqrt{5} \leq \gamma \leq 1/2$, $r_\text{MSCO}^{(-)}$, which is closer to the center than $r_\text{MSCO}^{(+)}$, corresponds to the ISCO.

From the above, the regions in which the stable circular orbits can exist are summarized as follows:
\begin{align}
    r_\text{MSCO}^{(+)} < r_0 \quad &\text{for}\ \ \gamma \geq \frac{1}{2}, \label{eq:con_stable_1} \\[5pt]
    \frac{2M}{\gamma} < r_0 < r_\text{MSCO}^{(-)} \ \ \text{and} \ \ r_\text{MSCO}^{(+)} < r_0 \quad &\text{for} \ \ \frac{1}{\sqrt{5}} \leq \gamma \leq \frac{1}{2}, \label{eq:con_stable_2} \\[5pt]
    r_0 > \frac{2M}{\gamma} \quad &\text{for} \ \ 0 < \gamma \leq \frac{1}{\sqrt{5}}.
\end{align}
For $\gamma \geq 1/2$, $r_\text{MSCO}^{(+)}$ corresponds to the ISCO. Taking the circular orbit condition~\eqref{conditions_co1} into account, we can find that unstable circular orbits can exist inside this ISCO, and further inside, there is a region where no unstable circular orbits can exist. For a more detailed discussion of the circular orbits in the ZV spacetime, see e.g.~\cite{Toshmatov:2019qih}. Similar calculations for circular orbits and epicyclic frequencies in the generalized ZV spacetime with a scalar field have been presented in~\cite{Azizallahi:2023rrv}.

\subsubsection{Derivation of an exact formula}
From Eq.~\eqref{eq:qc_freq}, the radial frequency of the quasi-circular orbit in terms of the proper time, as calculated in~\cite{Toshmatov:2019qih}, is given by
\begin{align}
    \omega_r &:= \sqrt{V^{\prime \prime}(r_0)} \\[8pt]
    &= \sqrt{\frac{M \left[ 1 - 2 (3+1/\gamma) M/r_0 + 2(1+1/\gamma)(2+1/\gamma) (M/r_0)^2 \right] }{ r_0^{3} \left[ 1 - (2+1/\gamma)M/r_0 \right] \left[ 1-M/(\gamma r_0) \right]^{2(1 - \gamma^2 )} \left[ 1-2M/(\gamma r_0) \right]^{1-\gamma+\gamma^2}}}. \label{eq:omega_r_zv}
\end{align}
Besides, the orbital angular velocity of the particle in terms of the proper time is given as follows~\cite{Toshmatov:2019qih}:
\begin{align}
    \omega_\phi := \dot{\phi} = \sqrt{\frac{M}{ r_0^3 \left[1-(2+1/\gamma)M/r_0 \right] \left[ 1-2M/(\gamma r_0) \right]^{1-\gamma} }}, \label{eq:omega_phi_zv}
\end{align}
where we have used Eqs.~\eqref{tdot_phidot} and \eqref{circular_E_L}. As can be seen from Eqs.~\eqref{eq:omega_r_zv} and~\eqref{eq:omega_phi_zv}, the orbital angular and radial frequencies generally do not coincide, and this causes the periapsis shifts of quasi-circular orbits as mentioned above. Note that if $M/r_{0}$ is sufficiently small, we have $\omega_r \approx \omega_\phi \approx \sqrt{M/r_0^3}$, which is consistent with the Newtonian result.

Substituting Eqs.~\eqref{eq:omega_r_zv} and~\eqref{eq:omega_phi_zv} into Eq.~\eqref{eq:qc_peri_pn_kousiki}, an exact formula for the periapsis shift of the quasi-circular orbit per round can be obtained as
\begin{align}
    \Delta \phi_\text{ZV,qc} &= 2 \pi \left\{ \sqrt{ \frac{ \left[ 1-M/(\gamma r_0) \right]^{2(1-\gamma^2)} \left[ 1 - 2 M/(\gamma r_0) \right]^{\gamma^2} }{ 1 - 2(3 + 1/\gamma)M/r_0 + 2 (1+1/\gamma)(2+1/\gamma) (M/r_0)^2 }} -1 \right\}. \label{peri_qc}
\end{align}
Expanding Eq.~\eqref{peri_qc} in powers of $M/r_0$, we obtain the PN expansion formula for the periapsis shift of the quasi-circular orbit as
\begin{align}
    \Delta \phi_\text{ZV,qc} = \frac{6 \pi M}{r_0} + \frac{\pi \left(22 + 6/\gamma -1/\gamma^2 \right) M^2}{r_0^2} 
    + \frac{\pi  \left(96 + 44/\gamma -3/\gamma^2 -2/\gamma^3 \right) M^3}{r_0^3}
    + \cdots. \label{peri_qc_PN}
\end{align}
The formula with even higher PN order terms can be obtained by taking $e \rightarrow 0$ in Eq.~\eqref{ZV_peri_PN_high} in Appendix~\ref{ZV_PN_high_order}. Note that Eqs.~\eqref{peri_qc} and~\eqref{peri_qc_PN} recover the formulae in the Schwarzschild spacetime~\cite{Einstein:1916vd,Schmidt:2008qi} for $\gamma=1$:
\begin{align}
    \Delta \phi_\text{S,qc} = 2\pi \left( \frac{1}{\sqrt{1-6M/r_0}} -1 \right) = \frac{6 \pi M}{r_0} + \frac{27 \pi M^2}{r_0^2} + \frac{135 \pi M^3}{r_0^3}
    +\cdots,
\end{align}
and in the CC spacetime~\cite{Bini:2005dy,Katsumata:2024qzv} for $\gamma \rightarrow \infty$:
\begin{align}
    \Delta \phi_\text{CC,qc} = 2 \pi \left[\frac{\mathrm{e}^{- \frac{1}{2} (M/r_0)^2}}{\sqrt{1 - 6 M/r_0 + 4 (M/r_0)^2}} - 1 \right] = \frac{6 \pi M}{r_0} + \frac{22 \pi M^2}{r_0^2} + \frac{96 \pi M^3}{r_0^3} + \cdots.
\end{align}

\renewcommand{\theequation}{D.\arabic{equation}}
\makeatletter
\@addtoreset{equation}{section}
\makeatother
\section{Concrete forms of expansion coefficients $Y_n(\chi)$} \label{Yn_forms}
In Sec.~\ref{sec:another_series}, the concrete forms of the expansion coefficients $Y_n(\chi)$ ($n=0, 1, 2, \ldots$) have been omitted. Here, we show their complete forms:
\begin{align}
    Y_0(\chi) &:= 1, \\[8pt]
    Y_1(\chi) &:= \frac{1}{3} \Bigl[ 3/\gamma + (1 -2/\gamma) (11 + 4/\gamma) M/\bar{p} + 2 \left( 3 -5 /\gamma + 9/\gamma^2 + 5/\gamma^3 \right) (M/\bar{p})^2 \notag \\
    &\quad -2 \left( 6 + 3/\gamma -4/\gamma^2 + 1/\gamma^4 \right) (M/\bar{p})^3 \Bigr] \sin \chi, \\[8pt]
    Y_2(\chi) &:= \frac{1}{6} \biggl\{ 2 (1 - 1/\gamma) \left[ 1 - M/(\gamma \bar{p}) \right]^2 \Bigl[ -3 (2 - 1/\gamma) + \left( 45 - 11/\gamma - 8/\gamma^2 \right) M/\bar{p} \notag \\
    &\quad - \left( 80 +11/\gamma - 33/\gamma^2 -12/\gamma^3 \right) (M/\bar{p})^2 \notag \\
    &\quad + 4 \left(12 + 8/\gamma - 13 /\gamma^2 -11/\gamma^3 - 2/\gamma^4 \right) (M/\bar{p})^3 \notag \\
    &\quad -2 (1 -2/\gamma) \left(2 + 3 /\gamma + 1/\gamma^2 \right)^2 (M/\bar{p})^4 \Bigr] \notag \\
    &\quad + \Bigl[ - 3 \left( 1 -4/\gamma^2 \right) + 2 \left( 33 + 31 /\gamma - 75 /\gamma^2 -34/\gamma^3 \right) M/\bar{p}  \notag \\
    &\quad - \left(169 +570 /\gamma - 270 /\gamma^2 - 654 /\gamma^3 - 184/\gamma^4 \right) (M/\bar{p})^2 \notag \\
    &\quad + 2 \left(126 + 330/\gamma + 465/\gamma^2 - 556/\gamma^3 - 621 /\gamma^4 -140/\gamma^5 \right) (M/\bar{p})^3 \notag \\
    &\quad -2 \left(70 + 186/\gamma + 346 /\gamma^2 + 255/\gamma^3 - 624 /\gamma^4 - 579 /\gamma^4 -122/\gamma^6 \right) (M/\bar{p})^4 \notag \\
    &\quad -8 \left(6 - 11/\gamma - 3 /\gamma^2 - 22/\gamma^3 - 24 /\gamma^4 + 56 /\gamma^5 + 57 /\gamma^6 + 13/\gamma^7 \right) (M/\bar{p})^5 \notag \\
    &\quad + 4 \left(2 + 3/\gamma +1/\gamma^2 \right)^2 \left(3 -6 /\gamma +10/\gamma^2 -9/\gamma^3 +3/\gamma^4 \right) (M/\bar{p})^6 \Bigr] \sin^2 \chi \biggr\},
\end{align}
and so forth. Note that they recover the coefficients in the CC case, $G_n(\chi)$, (see Eqs.~(3.26) and (3.27) in~\cite{Katsumata:2024qzv}) in the limit of $\gamma \rightarrow \infty$.

\renewcommand{\theequation}{E.\arabic{equation}}
\makeatletter
\@addtoreset{equation}{section}
\makeatother
\section{Brief review of the Zipoy-Voorhees spacetime} \label{app:review_ZV}
In this appendix, we briefly review the ZV spacetime. Axisymmetric, static, and vacuum solutions of Einstein's field equations can be described by the Weyl metric~\cite{Weyl:1917,Weyl:2012} (see also~\cite{Griffiths:2009dfa}). Using cylindrical coordinates $(t,\rho,z,\phi)$ with $-\infty < t < \infty$, $\rho \geq 0$, $-\infty < z < \infty$, and $0 \leq \phi < 2 \pi$, the line element is written as
\begin{align}
  \dd{s}^2 = - \mathrm{e}^{2 \Phi} \dd{t}^2 + \mathrm{e}^{2(\Lambda - \Phi)} ( \dd{\rho}^2 + \dd{z}^2 ) + \rho^2 \mathrm{e}^{-2 \Phi} \dd{\phi}^2, \label{weyl_met}
\end{align}
where the metric functions $\Phi$ and $\Lambda$ depend on $\rho$ and $z$. The vacuum Einstein's field equations imply
\begin{align}
    \pdv[2]{\Phi}{\rho} + \frac{1}{\rho} \pdv{\Phi}{\rho} + \pdv[2]{\Phi}{z} = 0, \quad \pdv{\Lambda}{\rho} - \rho \left[ \left(\pdv{\Phi}{\rho} \right)^2 - \left(\pdv{\Phi}{z} \right)^2 \right] = 0, \quad \pdv{\Lambda}{z} - 2 \rho \pdv{\Phi}{\rho} \pdv{\Phi}{z} = 0. \label{Weyl_eqs}
\end{align}
The first equation can be recognized as the simple three-dimensional Laplace equation in cylindrical coordinates. Therefore, $\Phi$ can be regarded as the analogy of the Newtonian potential. Once we find a solution for $\Phi$, we can also find a solution for $\Lambda$ by integrating the second and third equations. Note that the Minkowski spacetime is obviously obtained when $\Phi=0$ and $\Lambda=0$. 

The asymptotically flat solutions of the first equation for $\Phi$ can be expressed by multipole expansions (see e.g.~\cite{Griffiths:2009dfa}):
\begin{align}
    \Phi = - \sum_{n=0}^\infty \frac{a_n}{(\rho^2+z^2)^{(n+1)/2}} P_n \left( \frac{z}{\sqrt{\rho^2+z^2}} \right), \label{eq:sol_Phi}
\end{align}
where $P_n$ are Legendre polynomials. It is noteworthy that the coefficients $a_n$ in Eq.~\eqref{eq:sol_Phi} do not directly coincide with the actual mass multipole moments of a physical source. A calculation of the relativistic multipole moments $M_n$ is given in Appendix~\ref{app:multipole_moments}. The expression for $\Lambda$ corresponding to the solution~\eqref{eq:sol_Phi} is given as (e.g.~\cite{Griffiths:2009dfa})
\begin{align}
    \Lambda = - \sum_{i=0}^\infty \sum_{j=0}^\infty a_i a_j \frac{(i+1)(j+1)}{i+j+2} \frac{P_i P_j - P_{i+1} P_{j+1}}{(\rho^2+z^2)^{(i+j+2)/2}}.
\end{align}
The simplest asymptotically flat Weyl solution is generated by the spherically symmetric Newtonian potential of the point mass with mass $M$:
\begin{align}
    \Phi = - \frac{M}{R}, \quad  \Lambda = - \frac{M^2 \rho^2}{2 R^4}, \quad  R:= \sqrt{\rho^2 + z^2}. \label{cc_sol}
\end{align}
This is the CC solution~\cite{Chazy:1924,Curzon:1925}, which corresponds to putting the coefficients $a_n$ as $a_0=M$ and $a_n=0$ (for $n \geq 1$). Note that although the Newtonian potential that generates this spacetime is spherically symmetric, the spacetime itself is not spherically symmetric.

The ZV solution~\cite{Bach:1922,Darmois:1927,Zipoy:1966btu,Voorhees:1970ywo} is generated by a Newtonian potential sourced by a rod with the length $2l$ and the mass $M$, located at $\rho=0$ and $-l < z < l$. The corresponding potential is given by
\begin{align}
    \Phi = \sigma \log \left( \frac{R_- + z - l}{R_+ + z + l} \right), \quad R_{\pm}:=\sqrt{\rho^2+(z\pm l)^2},
\end{align}
where we have introduced the mass linear density $\sigma=M/(2l)$. Using the parameters $M$ and $l$, the solution for the metric~\eqref{weyl_met} can be expressed as
\begin{align}
    \mathrm{e}^{2\Phi} = \left(\frac{R_+ + R_- - 2l}{R_+ + R_- + 2l} \right)^{\gamma}, \quad \mathrm{e}^{2\Lambda} = \left[ \frac{(R_+ + R_-)^2 - 4 l^2 }{4 R_+ R_-} \right]^{\gamma^2}, \label{eq:sol_ZV}
\end{align}
where we have introduced a dimensionless parameter $\gamma$, known as the deformation parameter:
\begin{align}
    \gamma := \frac{M}{l} = 2 \sigma.
\end{align}
Note that the Minkowski spacetime is recovered for $\gamma \rightarrow 0$ (i.e. $M \rightarrow 0$ or $l \rightarrow \infty$). It is also noteworthy that the solution~\eqref{eq:sol_ZV} is invariant under the transformation $l \rightarrow -l$. By virtue of this, we can assume $l>0$ (therefore $\gamma > 0$) without loss of generality. As we will see in Appendix~\ref{app:multipole_moments}, this parameter $\gamma$ is related to the mass quadrupole moment and characterizes the deformation of the source from spherical symmetry. As for $\gamma >1$ ($0 < \gamma < 1$), the spacetime is oblate (prolate). Note that $\gamma$ is often denoted as $\delta$, and then the ZV spacetime is called the $\delta$-metric in some literature. The coefficients $a_n$ corresponding to the ZV solution are given by $a_n=0$ (for odd $n$) and $a_n=M l^n /(n+1)$ (for even $n$). When the mass linear density of the rod is $\sigma=1/2$ (i.e. $\gamma = 1$), Eq.~\eqref{eq:sol_ZV} recovers the Schwarzschild solution. It also recovers the CC solution~\eqref{cc_sol} in the limit $l \rightarrow 0$ (i.e. $\gamma \rightarrow \infty$), which corresponds to the case where the length of the rod that creates the analogical Newtonian potential, which generates the ZV spacetime, is zero.

Finally, we introduce the metrics of the ZV spacetime written in other coordinates. In terms of prolate spheroidal coordinates ($x$,$y$) such that
\begin{align}
    \rho = l \sqrt{(x^2-1)(1-y^2)}, \quad z = l x y, \label{eq:pro_sp_co}
\end{align}
the metric functions of the ZV spacetime~\eqref{eq:sol_ZV} is rewritten as
\begin{align}
    \mathrm{e}^{2 \Phi} = \left( \frac{x-1}{x+1} \right)^\gamma, \quad \mathrm{e}^{2\Lambda} = \left( \frac{x^2-1}{x^2-y^2} \right)^{\gamma^2},
\end{align}
and the metric~\eqref{weyl_met} becomes
\begin{align}
    \dd{s}^2 = -\mathrm{e}^{2\Phi} \dd{t}^2 + \Sigma^2 \left( \frac{\dd{x}^2}{x^2-1} + \frac{\dd{y}^2}{1-y^2} \right) + R^2 \dd{\phi}^2, \label{eq:met_zv_sp_co}
\end{align}
where
\begin{align}
    \Sigma^2 &= l^2 (x+1)^{\gamma(\gamma+1)} (x-1)^{\gamma(\gamma-1)} (x^2-y^2)^{1-\gamma^2}, \\[8pt]
    R^2 &= l^2 (x+1)^{1+\gamma} (x-1)^{1-\gamma} (1-y^2).
\end{align}
Furthermore, let us introduce the Erez-Rosen coordinates ($r$,$\theta$)~\cite{Erez:1959}
\begin{align}
    x = \frac{r}{l}-1, \quad y = \cos \theta. \label{eq:er_co}
\end{align}
As a side note, from Eqs.~\eqref{eq:pro_sp_co} and~\eqref{eq:er_co}, in the equatorial plane, the direct transformation from the Weyl radial coordinates $\rho$ to the Erez-Rosen radial coordinate $r$ is given as
\begin{align}
    \rho = r \sqrt{1 - \frac{2M}{\gamma r}}.
\end{align}
Notice that $\rho=r$ for the CC limit ($\gamma \rightarrow \infty$). Using Erez-Rosen coordinates~\eqref {eq:er_co}, the metric~\eqref{eq:met_zv_sp_co} can be rewritten as
\begin{align}
    \dd{s}^2 = -f^\gamma \dd{t}^2 + f^{\gamma(\gamma-1)} g^{1-\gamma^2} \left( f^{-1} \dd{r}^2 + r^2 \dd{\theta}^2 \right) + f^{1-\gamma} r^2 \sin^2 \theta \dd{\phi}^2, \label{eq:met_gamma}
\end{align}
where
\begin{align}
    f = 1 - \frac{2M}{\gamma r}, \quad g = 1 - \frac{2M}{\gamma r} + \frac{M^2 \sin^2 \theta}{\gamma^2 r^2}. \label{eq:metfunc_gamma}
\end{align}
The ZV metric written as in Eqs.~\eqref{eq:met_gamma} and~\eqref{eq:metfunc_gamma} is often called the $\gamma$-metric or the $\delta$-metric. Notice that it is obvious that the metric~\eqref{eq:met_gamma} exactly recovers the standard Schwarzschild metric for $\gamma=1$. This indicates that the Erez-Rosen radial coordinate $r$ coincides with the Schwarzschild radial coordinate (circumferential radius) for $\gamma=1$ as mentioned several times in the main text of this paper. We can also confirm that the metric~\eqref{eq:met_gamma} recovers the CC metric in the spherical coordinates for $\gamma \rightarrow \infty$:
\begin{align}
    \dd{s}^2 = - \mathrm{e}^{2\Phi} \dd{t}^2 + \mathrm{e}^{2(\Lambda-\Phi)} \left( \dd{r}^2 + r^2 \dd{\theta}^2 \right) + \mathrm{e}^{-2\Phi} r^2 \sin^2 \theta  \dd{\phi}^2,
\end{align}
where
\begin{align}
    \Phi = -\frac{M}{r}, \quad \Lambda = -\frac{M^2 \sin^2 \theta}{2 r^2}.
\end{align}
Note that this form of the CC solution can also be obtained by rewriting the CC solution written in Eqs.~\eqref{weyl_met} and \eqref{cc_sol} using variable transformation $\rho=r\sin \theta$ and $z=\rho \cos \theta$. In addition, focusing on the deviation from the spherically symmetric case ($\gamma=1$) and introducing the new dimensionless parameter $q$ as $\gamma=1+q$, Eq.~\eqref{eq:met_gamma} can be rewritten as
\begin{align}
    \dd{s}^2 = -f^{1+q} \dd{t}^2 + f^{-q} \left[ \left( 1 + \frac{M^2 \sin^2 \theta}{\gamma^2 r^2} f^{-1} \right)^{-q(2+q)} \left( f^{-1} \dd{r}^2 + r^2 \dd{\theta}^2 \right) + r^2 \sin^2 \theta \dd{\phi}^2 \right]. 
\end{align}
This form of the ZV metric is usually called the $q$-metric. Although the solution~\eqref{eq:sol_ZV} is now generally called the ZV metric, $\gamma$-metric, $q$-metric, and $\delta$-metric, it has been initially discovered by Bach and Weyl~\cite{Bach:1922}, and Darmois~\cite{Darmois:1927}. Therefore, to be more precise, we might call this spacetime the Bach-Weyl-Darmois-Zipoy-Voorhees (BWDZV) spacetime.

\renewcommand{\theequation}{F.\arabic{equation}}
\makeatletter
\@addtoreset{equation}{section}
\makeatother
\section{Multipole moments in the Zipoy-Voorhees spacetime} \label{app:multipole_moments}
The relativistic and coordinate-invariant definitions of multipole moments in general relativity have been proposed by Geroch~\cite{Geroch:1970cd}, Hansen~\cite{Hansen:1974zz}, Thorne~\cite{Thorne:1980ru}, and Beig and Simon~\cite{Beig:1981}. Algorithms to calculate the multipole moments have also been developed, for example, by Fodor, Hoenselaers, and Perj\'{e}s~\cite{Fodor:1989}. In this appendix, we present the calculation of the multipole moments in the ZV spacetime using a method proposed by Hoenselaers~\cite{Hoenselaers:1986gcr} and summarized more practically by Quevedo~\cite{Quevedo:1989rfm}.
 
Let us start by considering the Ernst potentials~\cite{Ernst:1967by}, $E$ and $\xi$, in the ZV spacetime:
\begin{align}
    E = \mathrm{e}^{2\Phi} = \left( \frac{x-1}{x+1} \right)^\gamma, \quad \xi = \frac{1-E}{1+E},
\end{align}
where we have used the prolate spheroidal coordinates. Introducing the inverse Weyl coordinate $\tilde{z}$ as
\begin{align}
    \tilde{z} = \frac{1}{z} = \frac{1}{l x y},
\end{align}
the Ernst potentials are  rewritten as
\begin{align}
    E(\tilde{z},y) = \left( \frac{1-l y \tilde{z}}{1+l y \tilde{z}} \right)^\gamma, \quad \xi(\tilde{z},y) = \frac{1-E(\tilde{z},y)}{1+E(\tilde{z},y)}. \label{Ernst_potentials}
\end{align}

Hoenselaers showed that, particularly in the static case, the mass multipole moments $M_n$ can be calculated by
\begin{align}
    M_n = m_n + d_n \quad (n=0,1,2,\ldots), \label{eq:multipole_cal}
\end{align}
where the first term $m_n$ can be obtained
\begin{align}
    m_n = \left. -\frac{h_{n+1}}{(n+1)!} \right|_{y=1, \tilde{z}=0}, \label{eq:m_n}
\end{align}
with the simple recurrence formula for $h_n$:
\begin{align}
    h_1 = \frac{1}{2 E} \dv{E}{\tilde{z}}, \quad h_n = \dv{h_{n-1}}{\tilde{z}} + 2 \xi h_1 h_{n-1} \quad (\text{for} \ \, n \geq 2), \label{recurrence_formula}
\end{align}
while the second term $d_n$ in Eq.~\eqref{eq:multipole_cal} is determined by comparing Eq.~\eqref{eq:multipole_cal} with the original Geroch-Hansen definition (see also~\cite{Fodor:1989}). It can be written in terms of $m_n$, for example,
\begin{align}
    d_0 &= d_1 = d_2 = d_3 = 0, \quad d_4 = -\frac{1}{7} A_{2,0} m_0, \quad d_5 = -\frac{1}{21} A_{2,0} m_1 - \frac{1}{3} A_{3,0} m_0, \label{eq:def_d_1} \\[8pt]
    d_6 &= \frac{1}{33} A_{2,0} m_0^{3} - \frac{5}{231} A_{2,0} m_2 - \frac{4}{33} A_{3,0} m_1 - \frac{8}{33} A_{3,1} m_0 - \frac{6}{11} A_{4,0} m_0, \label{eq:def_d_2}
\end{align}
where we have introduced the convenient notation
\begin{align}
    A_{i,j} := m_i m_j - m_{i-1} m_{j+1}.
\end{align}

Substituting the Ernst potentials $E$ and $\xi$ \eqref{Ernst_potentials} into Eq.~\eqref{recurrence_formula}, we can determine $h_n$. Then, using the obtained $h_n$ and Eq.~\eqref{eq:m_n}, $m_n$ up to $n=6$ can be obtained as follows:
\begin{align}
    m_0 &= l \gamma, \quad m_2 = \frac{l^3}{3} \gamma \left( 1-\gamma^2 \right), \quad
    m_4 = \frac{l^5}{15} \gamma \left( 3-5 \gamma^2 + 2 \gamma^4 \right), \\[8pt]
    m_6 &= \frac{l^7}{315} \gamma \left( 45 -98 \gamma^2 +70 \gamma^4 -17 \gamma^6 \right), \quad m_1=m_3=m_5=0.
\end{align}
Finally, using the obtained $m_n$ and Eqs.~\eqref{eq:multipole_cal},~\eqref{eq:def_d_1}, and~\eqref{eq:def_d_2}, we can get the relativistic multipole moments in the ZV spacetime up to $n=6$ (see e.g.~\cite{Herrera:2004fg}):
\begin{align}
    M_0 &= l \gamma, \label{eq:multipole_moment_m0} \\[8pt]
    M_2 &= \frac{l^3}{3} \gamma \left(1-\gamma^2\right), \\[8pt]
    M_4 &= \frac{l^5}{5} \gamma (1-\gamma^2) \left(1-\frac{19}{21} \gamma^2 \right), \\[8pt]
    M_6 &= \frac{l^7}{7}\gamma (1-\gamma^2) \left( 1 - \frac{292}{165} \gamma^2 + \frac{389}{495} \gamma^4 \right), \label{eq:multipole_moment_m6} \\[8pt]
    M_1 &= M_3 = M_5 = 0.
\end{align}
Introducing the ADM mass $M=l \gamma$, we can rewrite Eqs.~\eqref{eq:multipole_moment_m0}-\eqref{eq:multipole_moment_m6} as
\begin{align}
    M_0 &= M, \label{ZV_monopole} \\[8pt]
    M_2 &= -\frac{M^3}{3} \left( 1 - \frac{1}{\gamma^2} \right), \label{ZV_quadrupole} \\[8pt]
    M_4 &= \frac{M^5}{5} \left( 1 - \frac{1}{\gamma^2} \right) \left( \frac{19}{21} - \frac{1}{\gamma^2}  \right), \\[8pt]
    M_6 &= -\frac{M^7}{7} \left( 1 - \frac{1}{\gamma^2} \right) \left( \frac{389}{495} - \frac{292}{165} \frac{1}{\gamma^2} + \frac{1}{\gamma^4}  \right).
\end{align}
For the Schwarzschild case $\gamma=1$, all multipole moments except the monopole moment obviously vanish. Note that taking the CC limit $\gamma \rightarrow \infty$, they also reproduce the multipole moments in the CC spacetime (see e.g.~\cite{Herrera:2004fg}):
\begin{align}
    M_0 &= M, \quad M_2 = -\frac{1}{3} M^3, \quad M_4 = \frac{19}{105} M^5 , \quad M_6 = -\frac{389}{3465} M^7.
\end{align}

\bibliography{refs}
\bibliographystyle{JHEP}

\end{document}